\documentclass[showpacs,preprintnumbers,amsmath,amssymb,twocolumn,superscriptaddress,prb]{revtex4}
\usepackage{graphicx}
\usepackage{amsfonts}
\usepackage{amsmath, amsthm, amssymb,subfigure,dsfont}
\usepackage{dsfont}
\usepackage{lmodern}
\usepackage[usenames]{color}
\usepackage[usenames,dvipsnames]{xcolor}
\usepackage[normalem]{ulem}
\def\i{\mathrm{i}}     
\newcommand{\e}[1]{\mathrm{e}^{#1}}
\newcommand{\fat}[1]{\mathbf{#1}}

\newcommand{\ie}{i.e.}
\newcommand{\eg}{e.g.}

\begin{document}
\title{Three distinct types of quantum phase transitions in a (2+1)-dimensional array of dissipative Josephson junctions}
\author{Einar B. Stiansen}
\affiliation{Department of Physics, Norwegian University of
Science and Technology, N-7491 Trondheim, Norway}
\author{Iver Bakken Sperstad}
\affiliation{Department of Physics, Norwegian University of
Science and Technology, N-7491 Trondheim, Norway}
\author{Asle Sudb{\o}}
\affiliation{Department of Physics, Norwegian University of
Science and Technology, N-7491 Trondheim, Norway}

\date{Received \today}
\begin{abstract}
We have performed large-scale Monte Carlo simulations on a model describing a (2+1)-dimensional array of dissipative Josephson junctions. 
We find three distinct stable quantum phases of the system. The most ordered state features long-range spatial ordering in the phase $\theta$ of 
the superconducting order parameter, but temporal ordering only in spatial gradients $\Delta \theta$, not in $\theta$. Significantly, the most 
ordered state therefore does not have 3D $XY$ ordering. Rather, it features 2D spin waves coexisting with temporally disordered phases $\theta$. 
There is also an intermediate phase featuring quasi-long-range spatial order in $\theta$ coexisting with a gas of instantons in $\Delta \theta$. 
We briefly discuss possible experimental signatures of such a state, which may be viewed as a local metal and a global superconductor. The most 
disordered state has phase disorder in all spatio-temporal directions, and 
may be characterized as a gas of proliferated vortices coexisting with a gas of $\Delta \theta$-instantons. The phase transitions between these phases 
are discussed. The transition from the most ordered state to the intermediate state is driven by proliferation of instantons in $\Delta \theta$. The 
transition from the intermediate state to the most disordered state is driven by the proliferation of spatial point vortices in the background of a 
proliferated $\Delta \theta$-instanton gas, and constitutes a Berezinskii-Kosterlitz-Thouless phase transition. The model also features a direct phase 
transition from the most ordered state to the most disordered state, and this transition is neither in the 2D $XY$ nor in the 3D $XY$ universality 
class. It comes about via a simultaneous proliferation of point vortices in two spatial dimensions and $\Delta \theta$-instantons, with a complicated 
interplay between them. The results are compared to, and differ in a fundamental way from, the results that are found in dissipative quantum rotor 
systems. The difference originates with the difference in the values that the fundamental degrees of freedom can take in the latter systems compared 
to dissipative Josephson junction arrays.  
\end{abstract}	

%05.30.Rt 	Quantum phase transitions
%74.40.Kb 	Quantum critical phenomena
%74.50.+r 	Tunneling phenomena; Josephson effects
%74.81.Fa 	Josephson junction arrays and wire networks
\maketitle

\section{Introduction}
In general, dissipation suppresses quantum fluctuations and may support states of 
spontaneously broken symmetry. A remarkable consequence of this is the dissipation-driven 
quantum phase transition in a single resistively shunted Josephson junction in which the 
phase difference is localized in a minimum of the periodic Josephson potential.\cite{PhysRevLett.51.1506}
In the parameter space of Josephson coupling and dissipation strength, this corresponds physically to a 
phase diagram with one metallic phase and one superconducting phase. While the behavior of a single 
dissipative Josephson junction is theoretically well understood, the picture is less complete for spatially 
extended systems. Other than the fully disordered phase and the fully ordered phase expected from the single-junction 
system, the phase diagram of arrays of dissipative Josephson junctions is conjectured to host additional 
phases in both one \cite{0295-5075-9-5-003,Fazio2001235,PhysRevB.75.014522,Korshunov,Goswami-Chakravarty_JJ_array,PhysRevB.41.4009} 
and two\cite{0295-5075-9-5-003,Panyukov1987325,springerlink:10.1007/BF00683713} dimensions. These new, exotic phases can 
broadly be characterized by having various combinations of global and/or local phase fluctuations or order.
 
Most of the analytical works on similar models have been based on mean-field analyses or perturbative renormalization group arguments. 
Since these approaches are valid in a limited region of the parameter space, in particular regions far away from phase 
transitions, a non-perturbative approach is of importance. Previous numerical work on models of dissipative Josephson junctions 
has mostly focused on lower-dimensional systems. The first Monte Carlo simulation of a single dissipative Josephson junction was 
presented in Ref.~\onlinecite{PhysRevB.65.104516}, where a fluctuation measure of the imaginary-time path of the phase difference 
was introduced to characterize the localization transition. Improved and extended results for the same model were later reported 
in Ref. ~\onlinecite{PhysRevLett.95.060201}. For one spatial dimension, Ref. ~\onlinecite{PhysRevB.45.2294} reported four physically 
distinct phases for a dissipative Josephson junction chain. This simulation was performed on a dual model and not directly on the 
phase degrees of freedom. A model for a (2+1)-dimensional [$(2+1)$D] dissipative Josephson junction arrays (JJA) has been 
treated numerically by Ref. ~\onlinecite{PhysRevLett.94.157001}. In essence, their results support the simplest 
scenario for a zero-temperature phase diagram,\cite{Chakravarty_dissipative_PT,PhysRevB.37.3283} with one phase with and 
another without spatio-temporal order. This is also what was found in a large-scale Monte Carlo simulation on
the dissipative (2+1)D $XY$ quantum rotor model.\cite{PhysRevB.84.180503}

Finally, our investigations are also motivated by a rather different physical system which can be described by a closely related model. 
In Ref. ~\onlinecite{Aji-Varma_orbital_currents_PRL}, a quantum $XY$ model with bond dissipation in two spatial dimensions was used to 
describe quantum critical fluctuations in cuprate high-$T_c$ superconductors. The principal result of analytical work on this model is 
that the dissipation-driven quantum critical point is local, in the sense that the fluctuation spectrum is frequency dependent but momentum 
independent.\cite{Aji-Varma_orbital_currents_PRL} Although the physical system we have in mind primarily is that of a Josephson junction array, 
we return to a discussion of the possibilities of local quantum criticality later in the paper.\cite{footnote_previously46}

The purpose of this paper is to numerically investigate the phase diagram of a specific model of a (2+1)-dimensional 
dissipative Josephson junction array. We pay special attention to the manifest anisotropy that exists between 
the spatial and temporal dimensions. To be specific, the fluctuations of the quantum paths of the phase gradients will be explicitly characterized in 
terms of roughening transitions, allowing us to consider the (temporal) localization transition separately from the onset of (spatial) 
phase coherence. In particular, we will identify a partially superconducting phase with spatial, but no temporal phase coherence. 
This corresponds to a dissipative JJA which may sustain a nonzero Josephson current, but where one nonetheless has voltage fluctuations over each junction. 
We investigate two phase transitions where the spatio-temporal aspects are well separated and can be characterized in terms of 
either a spatial vortex-antivortex unbinding, or proliferation of instanton-like defects. We also discuss a direct quantum phase transition 
from an ordered state to a disordered state involving simultaneous disordering in space and imaginary time. This corresponds to a quantum phase
transition on a dissipative JJA where one transitions from a state sustaining a Josephson current and allowing no voltage fluctuations to a normal 
state, but via an unusual quantum phase transition that is neither in the 2D $XY$ nor 3D $XY$ universality class. 

\subsection{Model}
\label{sec:model}

An array of Josephson junctions consists of superconducting islands arranged in a regular network. Separating the islands are tunnel 
junctions in which Cooper pairs are able to tunnel from one superconducting grain to the neighboring grain. The fundamental degrees 
of freedom are the phases of the superconducting order parameters residing on the grains. A classical two-dimensional JJA is 
described by the 2D $XY$ model 
\begin{align} \label{eq:2DXY}  
H = -K \sum_{\langle\mathbf{x},\mathbf{x'}\rangle}\cos(\theta_{\mathbf{x}} - \theta_{\mathbf{x'}}),
\end{align}
where the summation goes over nearest neighboring sites on a square lattice. $\theta_{\mathbf{x}}$ is the phase of the complex order parameter of the 
superconducting grain at position $\mathbf{x}$.
Although the $U(1)$ symmetry of the phase variables cannot be spontaneously broken in two dimensions at any nonzero temperature 
(implicit in the classical description), the system nevertheless undergoes a Berezinskii-Kosterlitz-Thouless (BKT) transition in which 
it develops quasi-long-range order (QLRO) with power-law-decaying correlation functions in the low-temperature regime. 
The low-temperature phase corresponds to a dipole phase where the vortices and anti-vortices of the phase field are bound in pairs. 
At the transition the vortices proliferate and destroy the QLRO. For a given phase configuration, a single vortex is identified on a 
plaquette by a nontrivial line integral of the phase difference around the plaquette, taking the compactness of the phase field into 
account.

The quantum generalized version of the model includes two additional terms describing  quantum fluctuations in imaginary time $\tau$. The action reads\cite{PhysRevB.73.064503,PhysRevB.37.3283,PhysRevB.45.2294,Fazio2001235,PhysRevLett.94.157001,Aji-Varma_orbital_currents_PRL,PhysRevB.82.174501}
\begin{align}\label{SBond}
&S =  \frac{1}{2E_C}\sum_{\fat{x}}\int_0^\beta \mathrm{d}\tau \left(\frac{\partial\theta_{\fat{x},\tau}}{\partial \tau}\right)^2 \\ \nonumber
				&- K   \sum_{\langle\fat{x},\fat{x'}\rangle}\int_0^\beta\mathrm{d}\tau \cos(\Delta\theta_{\fat{x},\fat{x'},\tau} ) \\ \nonumber
				 &+ \frac{\alpha}{2}\sum_{\langle\fat{x},\fat{x'}\rangle}\int_0^\beta\int_0^\beta \mathrm{d}\tau  \mathrm{d}\tau'\left(\frac{\pi}{\beta}\right)^2
				 \frac{(\Delta\theta_{\fat{x},\fat{x'},\tau} -\Delta\theta_{\fat{x},\fat{x'},\tau'})^2}{\sin^2(\frac{\pi}{\beta}|\tau-\tau'|)},
\end{align}
where we have defined the lattice gradient $\Delta\theta_{\fat{x},\fat{x'},\tau} = \theta_{\fat{x},\tau} - \theta_{\fat{x'},\tau}$. The first term describes 
the self-capacitance of a single island, the second term is the familiar Josephson interaction, coupling each superconducting island to the nearest neighbors 
by a periodic potential. The last term describes the ohmic dissipation as modeled by a bath of harmonic oscillators coupling to the bond variables.\cite{Caldeira-Leggett}

A subtle consequence of the presence of this ohmic shunt mechanism is that the phase variables become noncompact,\cite{Schön1990237} as the dissipation term 
in Eq. \eqref{SBond} breaks the $2\pi$-periodicity of the Josephson potential. Thus, the phases are no longer defined with compact support 
$\theta \in  [ - \pi , \pi \rangle$, as they would be in the non-dissipative case or in a (2+1)D dissipative quantum rotor model. Instead, 
we have $\theta \in \langle -\infty, \infty \rangle$. {\it The impact of this decompactification on the problem is enormous}. It reflects that a sudden increase 
along imaginary time in the phase difference, \eg, $\Delta\theta_{\mathbf{x},\mathbf{x'},\tau} \rightarrow \Delta\theta_{\mathbf{x},\mathbf{x'},\tau} + 2\pi$, 
would produce a voltage imbalance over the barrier. 
A dissipative, measurable current would then flow through the shunting resistors until the imbalance is relaxed. Hence, the variables cannot be 
defined modulo $2\pi$, since $\Delta\theta_{\mathbf{x},\mathbf{x'},\tau}$ and $\Delta\theta_{\mathbf{x},\mathbf{x'},\tau} + 2\pi$ represent distinguishable states. 
The noncompactness of the variables implies that we may no longer identify vortices in the same manner as described above, as a line integral around a plaquette 
always yields zero for a noncompact phase field. In App. \ref{sec:noncompact}, we introduce a reformulation of the phase variables in terms of a compact part and an 
additional field describing the tunneling between wells in the extended Josephson potential. This enables us to identify vortices in the compact part of the phase. 
The phase transitions involving spatial ordering may therefore still be described by vortex proliferation even though the variables are of a noncompact nature.

As a description of a dissipative JJA, there are a few simplifications built into the action \eqref{SBond}. We have only considered the effect of self-capacitance and 
neglected mutual capacitive coupling with neighboring grains. Also, the dissipation term only accounts for one source of dissipation, namely the flow of normal 
electrons through the shunting resistors. Additional dissipative effects like quasiparticle tunneling\cite{Schön1990237} and Cooper pair 
relaxation\cite{PhysRevB.68.214515,PhysRevB.75.014522} have been neglected.

In order to study the behavior of a two-dimensional array of Josephson junctions at zero temperature under the influence of ohmic dissipation, we perform large scale 
Monte Carlo simulations on a discretized version of Eq. \eqref{SBond},
 \begin{align}\label{SBond_Dis}
S =  &\frac{K_\tau}{2}\sum_{\fat{x}}^N\sum_\tau^{N_\tau} (\theta_{\fat{x},\tau+1}-\theta_{\fat{x},\tau})^2 \\ \nonumber
				- &K             \sum_{\langle\fat{x},\fat{x'}\rangle}\sum_\tau^{N_\tau} \cos(\Delta\theta_{\fat{x},\fat{x'},\tau} ) \\ \nonumber
				 + &\frac{\alpha}{2}\sum_{\langle\fat{x},\fat{x'}\rangle}\sum_{\tau\neq\tau'}^{N_\tau}\left(\frac{\pi}{N_\tau}\right)^2
				 \frac{(\Delta\theta_{\fat{x},\fat{x'},\tau} -\Delta\theta_{\fat{x},\fat{x'},\tau'})^2}{\sin^2(\frac{\pi}{N_\tau}|\tau-\tau'|)}.
\end{align}
Here, $K_\tau = 1/E_C\Delta\tau$ and the spatial coupling has been renamed $ K\Delta\tau \rightarrow K$. Our goal is to investigate the behavior of the system in 
the $K$-$\alpha$-space, $K_\tau$ therefore defines the energy scale and will be kept at suitable values in the simulations. The variables are defined on the vertices 
of a three-dimensional cubic grid. The spatial linear extent of the grid is given by $N$, and the number of Trotter slices used to discretize the temporal direction 
is given by $N_\tau$. Thus, $\Delta\tau = \beta/N_\tau$, and the size of the space-time lattice is $N \times N \times N_\tau$. Periodic boundary conditions in imaginary 
time are implicit from the path integral construction, and are also applied in the spatial directions in the standard manner. The noncompactness of the variables also 
dictates the form of the kinetic term. Because $\theta$ is an extended variable, its derivative must be expressed by discretized differentiation. We refer to the appendix of 
Ref. \onlinecite{PhysRevB.83.115134} for details.

\subsection{Outline and overview of main results}
\label{sec:overview}

\begin{figure}
\centering
\includegraphics[width=0.45\textwidth]{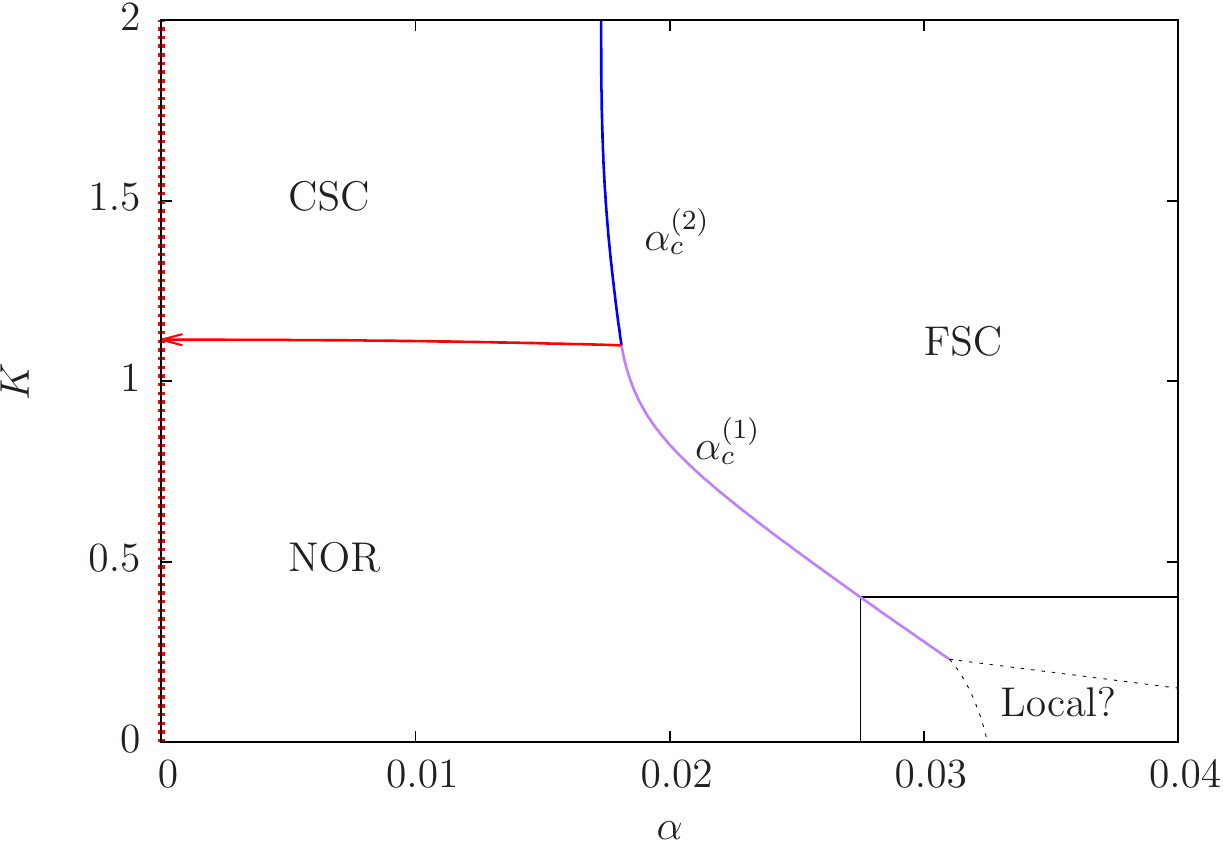}
\caption{A schematic phase diagram of the system defined by Eq. \ref{SBond_Dis}, based on the Monte Carlo calculations presented below. Here, we have used 
the value $K_\tau=0.002$, corresponding to the parameters in Sec. \ref{sec:result_strong}. NOR refers to the normal phase, where vortices are proliferated 
and the bonds $\Delta \theta$ are disordered in the $\tau$ direction. CSC refers to the critical superconducting state, where the $\theta$ variables feature 
power-law correlations in space, while $\Delta\theta$ remains disordered. FSC refers to the fully bond-ordered superconducting state, which features an 
additional ordering compared to the CSC phase, namely $\Delta \theta$-ordering in imaginary-time direction. A hypothetical fourth, local phase has \emph{not} been observed in our simulations, as indicated by the box in the lower right corner. See text in Sec. \ref{sec:overview} for more 
details.}
\label{fig:phasediag}
\end{figure}

For outlining the roadmap to this paper, the phase diagram of the system is helpful. This  is illustrated schematically in Fig. \ref{fig:phasediag}.
\emph{In all regions of the phase diagram, the phases $\theta$ are disordered in the imaginary-time direction.} 

In Sec. \ref{sec:observables}, we introduce the various observables used to identify the phases and phase transitions of the model defined in 
Eq. \eqref{SBond_Dis}. In   Sec. \ref{sec:MC}, the details of the Monte Carlo simulations are presented in a concise form.

In Sec. \ref{sec:result_strong}, we take a large value of the Josephson coupling $K$ and investigate the behavior of the system as it crosses from the CSC phase 
to the FSC phase in Fig. \ref{fig:phasediag} upon increasing $\alpha$. There is a phase transition at a critical dissipation strength, $\alpha_c^{(2)}$, 
above which the system is fully bond-ordered superconducting (FSC). For $\alpha<\alpha_c^{(2)}$ the system features unbounded temporal fluctuations, while at the  
same time featuring spatial phase coherence. Due to algebraically decaying spatial correlations in this regime, we will refer to the phase as critical superconducting 
(CSC). In other words, the phase configurations of the system rotate more or less as a ``rigid body'' in time, thus at the same time giving rise to a finite superfluid 
density (helicity modulus) as well as voltage fluctuations across the junctions. A detection of the CSC phase thus requires simultaneous measurements of the 
superfluid density of the system, as well as ac measurements of voltages across junctions. 

In Sec. \ref{sec:BKT}, we consider the transition between the NOR phase and the CSC phase, and this is found to be a purely spatial phase transition 
of the BKT type.

In Sec. \ref{sec:results_low}, we investigate the response of the system to increasing dissipation at low and intermediate Josephson couplings,
as it crosses from the NOR phase to the FSC phase in Fig. \ref{fig:phasediag}. This is the most difficult case to analyze, as the system transitions 
from a spatio-temporally disordered phase directly to the to the spatio-temporally ordered state FSC  upon crossing the critical line 
$\alpha_c^{(1)}$.

In Sec. \ref{sec:discussion}, the topological defects driving the various phase transitions as well as how such a model may exhibit local quantum criticality (LQC), are 
discussed. This may be briefly summarized as follows. 

On the line separating CSC from NOR, and on the line separating CSC from FSC, the spatial and temporal aspects of the phase transitions can be considered 
separately. The CSC--NOR transition is driven by point vortices and is in the 2D $XY$ universality class. The FSC--CSC transition is driven by instantons in $\Delta \theta$ 
and may be characterized as a roughening transition in the space of $\Delta \theta$. On the critical line $\alpha_c^{(1)}$, there is a complicated interplay between 
temporal and spatial fluctuations. This critical line is neither in the 2D $XY$ nor in the 3D $XY$ universality class. 

An additional fourth phase could conceivably have been present in the phase diagram, featuring temporal order and unbound vortices. The most likely 
position in the phase diagram for such a hypothetical phase would be at weak Josephson coupling and strong dissipation strength. 
This is shown by the dotted lines within the box in lower right corner of Fig. \ref{fig:phasediag}. The local transition line would involve ordering of 
temporal fluctuations without onset of spatial phase coherence, and as such would describe a local quantum critical point. Our simulations, however, 
show no sign of such behavior in the parameter range we have considered.  

The limit $\alpha=0$ is in principle ill-defined in this model since a finite dissipation is essential for the decompactification of the variables. This is indicated by 
drawing the $\alpha=0$ axis as a red dotted line in Fig. \ref{fig:phasediag}. Thus, the value $\alpha=0$ is also a singular endpoint of the horizontal (red) line in the 
phase diagram, and this is indicated by terminating this line in an arrow.

In App. \ref{sec:noncompact}, we provide some more details and discussion on the fundamental implications of the noncompactness of the phase field. In  
App. \ref{appendix1}, we take a closer look at the NOR phase and investigate the description of Refs. ~\onlinecite{PhysRevB.72.060505}, ~\onlinecite{PhysRevB.73.064503} 
of such a normal phase as a so-called floating phase.

\section{Observables}\label{sec:observables}

In order to describe the various phases and transitions introduced in the previous section, several quantities will be calculated. To monitor the degree of 
(spatial) superconducting order, we calculate the spatial helicity modulus, or phase stiffness. This quantity measures the increase in the free energy when 
applying an infinitesimal twist across the system, $\theta_{\fat{x}} \rightarrow \theta_{\fat{x}} - \boldsymbol\delta \cdot \fat{x}$. It probes the degree 
of phase coherence in the system and thus its ability to sustain a supercurrent. The only term in the action that contributes to the helicity modulus is the 
Josephson interaction term. Hence, the helicity modulus $\Upsilon_x$ is given by 
\begin{align}\label{helicity}
\Upsilon_x &= \frac{1}{N^2 N_\tau}\left\langle \sum_{\langle \fat{x},\fat{x'}\rangle}\sum_{\tau}^{N_\tau}\cos(\Delta\theta_{\fat{x,\fat{x'},\tau}})\right\rangle \\ \nonumber
&-\frac{K}{N^2 N_\tau}\left\langle \left( \sum_{\langle \fat{x},\fat{x'}\rangle}\sum_{\tau}^{N_\tau}\sin(\Delta\theta_{\fat{x},\fat{x'},\tau}) \right)^2 \right\rangle.
\end{align}
Here, the brackets indicate ensemble averaging.  In the context of the classical 2D $XY$ model, $\Upsilon_x =0$ defines the disordered state where vortices are 
proliferated. In the same manner, $\Upsilon_x\neq 0$ signals the finite rigidity of the quasi-ordered state.

The same $XY$ models used to describe superconducting systems also describe magnetic systems of planar spins, and the superconducting phase $\theta$ can formally 
be associated with the direction of the $XY$ spins. Conventionally, the order of a superconducting system is therefore often described by a magnetization order 
parameter
\begin{equation}\label{eq:magn}
	m = \frac{1}{N^2 N_\tau} \sum_{\fat{x},\tau}\e{\i \theta_{\fat{x},\tau}},
\end{equation}
which probes the uniformity of the spin direction across the entire (2+1)-dimensional volume of the system. 

It should be noted that these two order parameters are periodic and consequently insensitive to tunneling events where the phase difference on a single junction jumps 
to a neighboring potential well, $\Delta\theta \rightarrow \Delta\theta +2\pi$. Consequently, $\Upsilon_x$ and $m$  do not probe the dissipation-induced localization 
\emph{per se}. In order to quantify this, we calculate the mean square displacement (MSD) of the bond variable $\Delta\theta$ along imaginary time, 
\begin{align}\label{W}
W_{\Delta\theta}^2(N_\tau) = \frac{1}{N_\tau}\left\langle\sum_\tau^{N_\tau} \left(\Delta\theta_\tau - \overline{\Delta\theta} \right)^2 \right\rangle.
\end{align} 
Here, we have defined $\overline{\Delta\theta} = 1/N_\tau\sum_\tau \Delta\theta_\tau$. The MSD is often used in the context of stochastically growing interfaces 
or diffusion processes, and it is natural to adopt some concepts from these areas for our problem. For instance, the degree to which the imaginary-time history of 
$\Delta\theta$ may be regarded as ``rough'' can be quantified by the scaling characteristics of the MSD with the length $N_\tau$ of the ``interface''.  Normally, one finds
\begin{align}\label{w_scale}
W_{\Delta\theta}^2 \propto N_\tau^{2H}
\end{align} 
if the imaginary-time history of $\Delta\theta$ describes self-affine configurations. $H=1/2$ corresponds to a Markovian random walk, and such linear scaling of the 
MSD is also referred to as normal diffusion.  A deviation from linear growth of $W_{\Delta\theta}^2$ as a function of $N_\tau$ is the hallmark of \emph{anomalous diffusion}.\cite{Metzler20001}
In particular, $H<1/2$ is referred to as subdiffusive behavior. A smooth interface is characterized by the MSD being independent of 
the system length.

To describe the phases and phase transitions, we will also investigate correlations of the order parameter field considered in Eq. \eqref{eq:magn}. We define the spatial and temporal correlation function by
\begin{align}\label{eq:corrq}
G_{\theta}(\mu;q) = \left \langle \e{\mathrm{i}q(\theta_\mu-\theta_0)} \right \rangle,
\end{align}
where $\mu \in \{\fat{x},\tau\}$. The extra factor $q$ in the exponent is introduced for later reference in App. \ref{appendix1}, but will be set to the conventional value $q=1$ otherwise. In App. \ref{appendix1} we will also consider bond correlations, defined here for convenience as 
\begin{align}\label{eq:corrqDelta}
G_{\Delta\theta}(\mu;q) = \left \langle \e{\mathrm{i}q(\Delta\theta_\mu-\Delta\theta_0)} \right \rangle.
\end{align}

For completeness we also present the susceptibility of the action,
\begin{align}\label{susceptibility}
\chi_S = \frac{1}{N^2 N_\tau}\langle \left(S- \langle S \rangle \right)^2\rangle,
\end{align}
as an additional means of locating the expected dissipation-induced phase transitions. This is the quantum mechanical equivalent of the classical heat capacity and 
is expected to present a nonanalyticity at a critical point.

\section{Details of the Monte Carlo Calculations}\label{sec:MC}
Considerable progress has been made in constructing new, effective, non-local algorithms for 
long-range-interacting systems with extended variables.\cite{PhysRevLett.95.060201, 1742-5468-2005-12-P12003, Werner_dissipative_MC_algorithms} 
However, these algorithms are presently restricted to (0+1)D systems, and do not seem to generalize easily to 
$N > 1$\cite{1742-5468-2005-12-P12003}. In the Monte Carlo simulations, 
we have therefore combined local updates with a parallel tempering algorithm\cite{Hukushima_parallel_tempering,Katzgraber_MC} 
in which several systems are simulated simultaneously at different coupling strengths. 

A Monte Carlo sweep corresponds to proposing a local update by the Metropolis-Hastings algorithm sequentially for every grid point in 
the system. The proposed new phases are generated by first randomly choosing to increase or decrease the value, then propagating the 
value by a random increment of size $2\pi n/32$, where $n\in\{1,32\}$. In other words, the continuous symmetry of the variables is 
emulated by $32$ discrete states per $2\pi$ interval. We have confirmed that adding additional states will not change the results. 

After a fixed number of Monte Carlo sweeps a parallel tempering move is made. In this move, a swap of 
configurations between two neighboring coupling values is proposed, and the swap is accepted with probability $\Xi_{PT}$ 
given by 
\begin{equation}
 \Xi_{PT} =
\begin{cases}
1 &\mathrm{if} \ \Delta < 0, \\
\e{-\Delta} &\mathrm{if} \ \Delta \geq 0.
\end{cases}
\end{equation}
Here, $\Delta = \kappa'(\bar{S}[X;\kappa']- \bar{S}[X';\kappa']) - \kappa(\bar{S}[X;\kappa] - \bar{S}[X';\kappa]) $, where $\kappa$ 
is the coupling value varied, representing in our case $\alpha$ or $K$, and $X$ represents the phase configuration. $\bar{S}$ 
indicates the part of the action conjugate to the coupling parameter $\kappa$. Both the Metropolis updates and the parallel 
tempering swaps are ergodic and respect detailed balance.

All Monte Carlo simulations were initiated with a random configuration. Depending on system size, various numbers of sweeps were performed for each coupling value. Error bars are provided for all observables except correlation functions, but are usually smaller than the data points. Measurements on which we perform scaling have, broadly speaking, a relative error well below $1\%$. 
The Mersenne-Twister\cite{Mersenne_Twister} random number generator was used in all simulations and the random number generator on each CPU was independently seeded. 
It was confirmed that other random number generators yielded consistent results. In some simulations we also made use of the Ferrenberg-Swendsen reweighting technique,
\cite{PhysRevLett.63.1195} which enables us to continuously vary the coupling parameter after the simulations have been performed.

In order to identify sharply defined nonanalyticities and observe converged scaling of $W_{\Delta\theta}^2$ at the dissipation-induced phase transitions, relatively 
large values of $N_\tau$ are needed. This limits the range of spatial sizes accessible in simulations with a single-site update algorithm. In the sections where we 
focus on the temporal scaling, we have fixed the spatial size at $N=20$ and varied the temporal size in the range $N_\tau=50$ to $N_\tau=350$. In Sec. \ref{sec:BKT} 
we find that in the CSC phase the temporal size of the system is irrelevant in determining the spatial properties of the system. Consequently, the temporal size is 
fixed at $N_\tau=20$ and the spatial size is varied in the range $N=10$ to  $N=100$. To investigate the spatial correlations in $\theta$ across the NOR--FSC phase transition, we have also performed simulations on a $N_\tau=30$ system with $N=50$ and $N=100$.

\section{The CSC--FSC transition $\alpha_c^{(2)}$}\label{sec:result_strong}

In this section, we consider the behavior of the system under the influence of strong Josephson coupling $K$, i.e. for a 
value of $K$ where the corresponding classical system would be topologically ordered even in the absence of dissipation (above the horisontal line 
in the phase diagram of Fig. \ref{fig:phasediag}). The coupling parameter  will be fixed at $K=1.5$ in this section, while 
the dissipation strength $\alpha$ is varied. We will use a quantum coupling $K_\tau = 0.002$. The main focus is on scaling 
of observables describing temporal fluctuations. Hence, the spatial system size is fixed at $N=20$.

\begin{figure}
  \centering
   \includegraphics[width=0.45\textwidth]{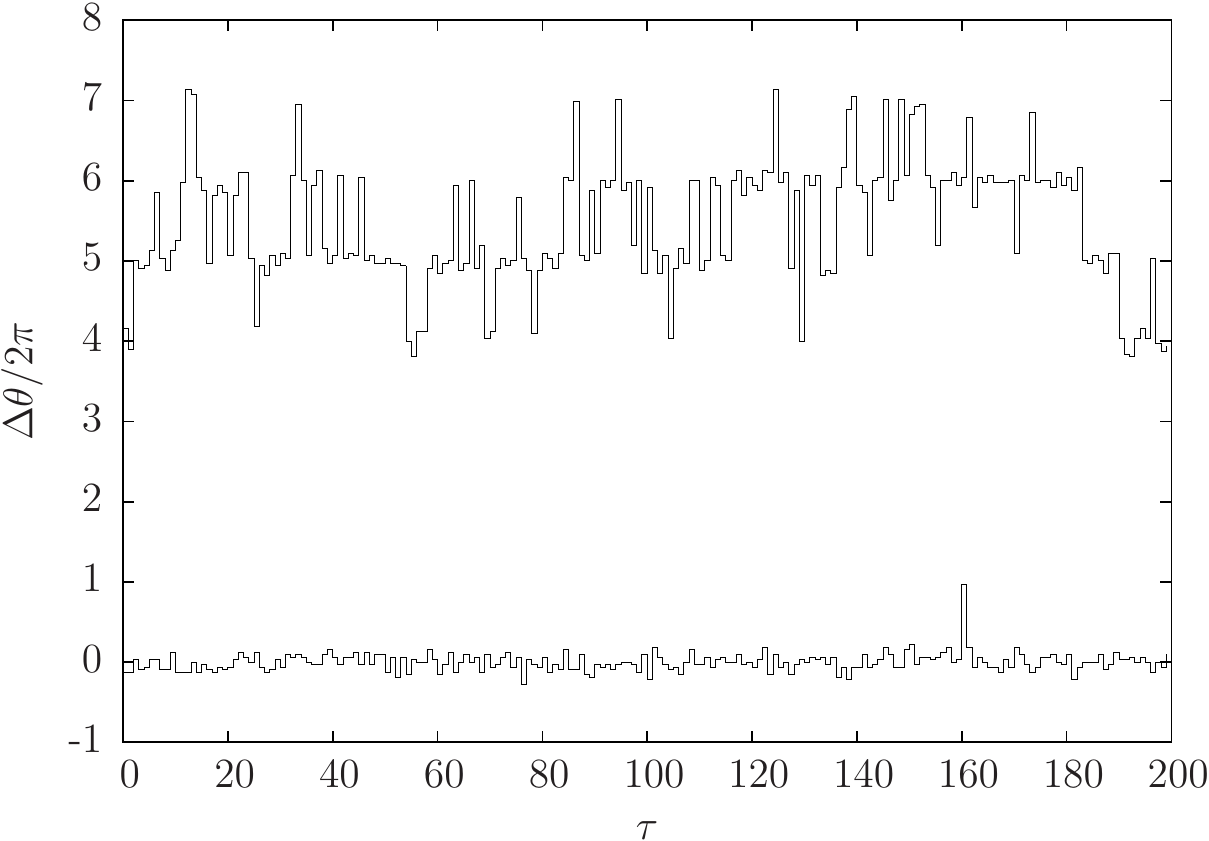}
  \caption{The bond variable $\Delta\theta$ as a function of imaginary time $\tau$ for two different values of dissipation strength, $\alpha=0.0102$ and $\alpha = 0.0281$, 
in the strong Josephson coupling regime. In the topmost curve, the bond variable clearly spends most of the time in the vicinity of the potential minima, although tunneling 
events between minima are frequent. The lowermost curve corresponds to the fully bond-ordered superconducting state where $\Delta\theta$ is localized and $W_{\Delta\theta}^2$ 
does not scale with $N_\tau$. Note that in the CSC phase (topmost curve) the quantum paths of $\Delta\theta$  are well described in terms of instantons where the fluctuations
in imaginary time are mostly given by integer multiples of $2 \pi$, in contrast to the situation 
for the corresponding quantum paths of $\Delta\theta$ in the NOR phase, see Fig. \ref{fig:BondRough}.}
  \label{fig:states_strong}
\end{figure}

We start by presenting typical configurations of the bond variable, $\Delta\theta$, as a function of $\tau$. At 
strong coupling, the bond variables are located predominantly in the vicinity of the potential minima located at $2\pi n$, where $n$ is an integer. Due to the 
noncompact nature of the variables, $\Delta\theta$ are free to tunnel between neighboring minima at weak dissipation. When considering the single-junction problem, 
this sudden tunneling of the bond variable from one Trotter slice to the next, $\Delta\theta_{\tau+1} - \Delta\theta_\tau \approx 2\pi n_\text{I}$, is often 
referred to as instanton or anti-instanton configurations, depending on the sign of the integer valued ``instanton charge'', $n_\text{I}$. Note that the noncompactness 
allows for tunneling of $\Delta\theta$ also between minima of the potential located further away than nearest neighbor. This corresponds to instanton charges with values 
larger than unity. The tunneling behavior is easily identified in the topmost curve in Fig. \ref{fig:states_strong}, where frequent instantons and anti-instantons are 
apparent. In this temporally disordered state, 
the quantum paths of $\Delta \theta$ appear to be well described in terms of a gas of proliferated instantons. Beyond a threshold value of $\alpha$, we observe a localization of $\Delta\theta$ in one of the minima of the Josephson potential.
The imaginary-time history of a bond variable corresponding to this phase forms an essentially smooth surface and is given in the lower curve in Fig. \ref{fig:states_strong}. 
However, even though the phase gradients are localized, closely bound pairs of instantons and anti-instantons may still be present. 
 
\begin{figure}
  \centering
   \includegraphics[width=0.45\textwidth]{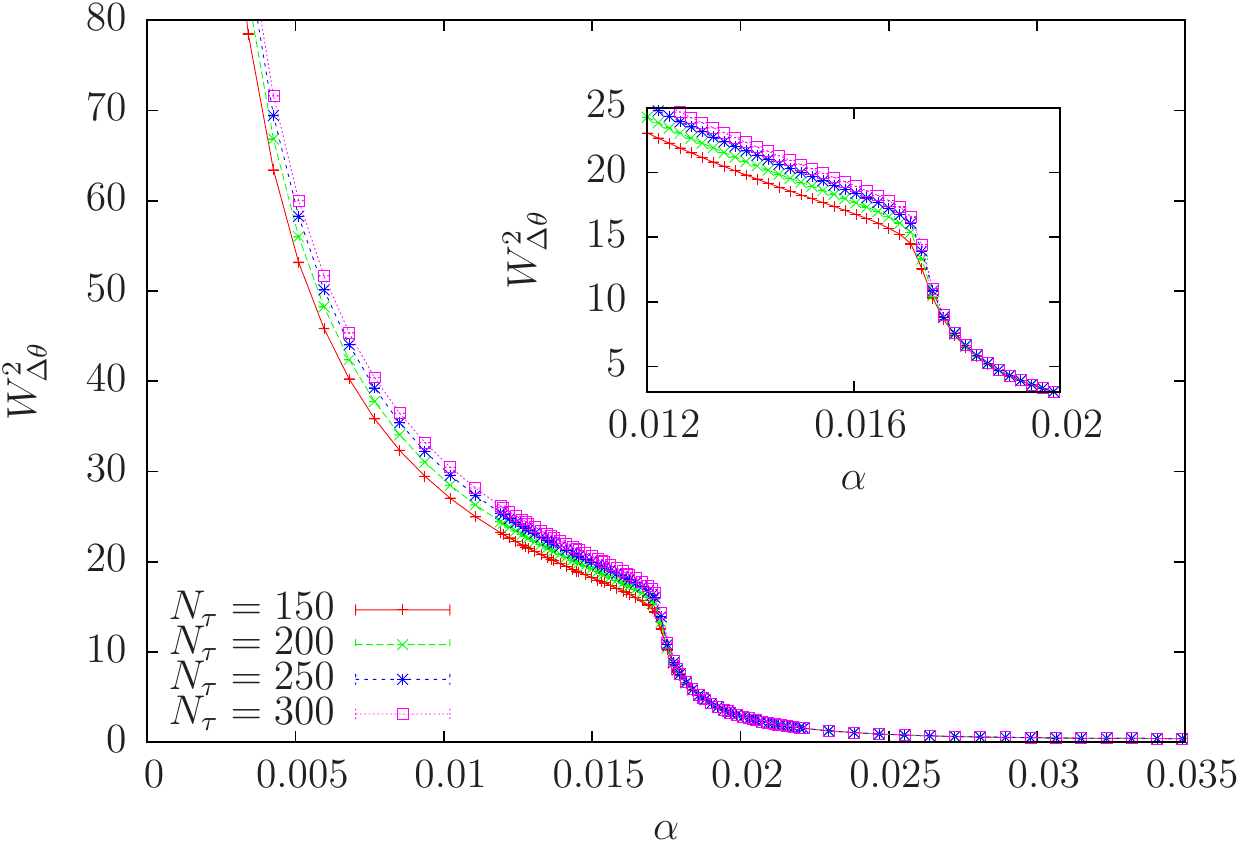}
  \caption{ (Color online) $W_{\Delta\theta}^2$, Eq. \eqref{W}, as a function of dissipation strength for a system with $K_\tau =0.002$, $K=1.5$, $N=20$ and various values of 
$N_\tau$. Note the kink in the curves at $\alpha=\alpha_c^{(2)}$ and the saturation of $W_{\Delta\theta}^2$ at a finite value for $\alpha> \alpha_c^{(2)}$. Error bars are smaller than the data points. Inset: A blow-up of 
the region around $\alpha_c^{(2)}$.   }
  \label{fig:rough_strong}
\end{figure}

In Fig. \ref{fig:rough_strong}, we show the mean square displacement as a function of dissipation strength. The temporal bond fluctuations are clearly 
suppressed for increasing values of $\alpha$. The different curves represent different values of $N_\tau$. Two regions of different scaling behavior 
of $W_{\Delta\theta}^2$ as a function of $N_\tau$ can immediately be discerned. For weak dissipation, $W_{\Delta\theta}^2$ increases with $N_\tau$, 
while $W_{\Delta\theta}^2$ is independent of the temporal size at strong dissipation. Separating the two regions is a precipitous drop in 
$W_{\Delta\theta}^2$ at a value of $\alpha$ that we will identify as the localization transition point $\alpha_c^{(2)}$.

\begin{figure}
  \centering
   \includegraphics[width=0.45\textwidth]{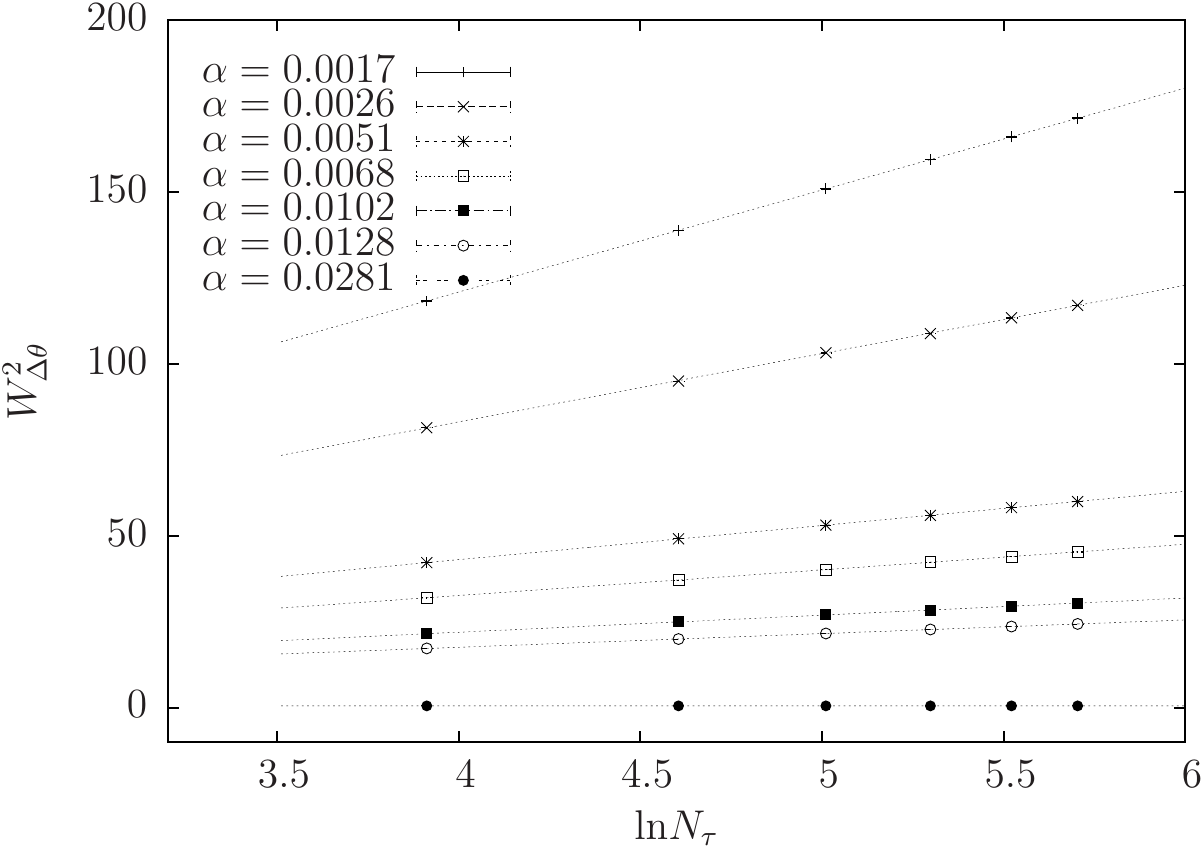}
  \caption{$W^2_{\Delta\theta}$ as a function of $\ln N_\tau$ for various values of the dissipation strength $\alpha$ ranging from the weak-dissipation 
    limit to the ordered state at the top and bottom, respectively. The dotted lines indicate the logarithmic growth of $W_{\Delta\theta}^2$. Error bars are much smaller than the data points.} 
  \label{fig:rough_scale_strong}
\end{figure}

Further information on the delocalized phase (CSC) can be found from investigating the dependence of $W^2_{\Delta\theta}$ on the temporal system 
size $N_\tau$. Here, the MSD scales with $N_\tau$ according to
\begin{align}\label{log_scale}
W^2_{\Delta\theta} = a(\alpha)\ln N_\tau,
\end{align}
where $a(\alpha)$ is a continuously varying proportionality constant. In Fig. 
\ref{fig:rough_scale_strong} we have plotted $W^2_{\Delta\theta}$ as a function of $\ln N_\tau$ for 
various dissipation strengths. All but the lowest curve represent dissipation strengths well below $\alpha_c^{(2)}$. 
A clear logarithmic dependence is seen for all values of dissipation strength in the CSC phase. The lowest curve with zero 
slope corresponds to $\alpha > \alpha_c^{(2)}$, where temporal fluctuations are effectively quenched and $W^2_{\Delta\theta}$ does not 
scale with $N_\tau$. In this way the increase of temporal fluctuations in $\Delta \theta_\tau$ as $\alpha$ is lowered may also be interpreted as a roughening transition 
at which the profile described by $\Delta\theta$ changes from smooth to rough. However, it should be noted that the logarithmic scaling presented 
in Fig. \ref{fig:rough_scale_strong} does not conform to the scaling ansatz \eqref{w_scale} for a self-affine interface. Instead, 
$\Delta\theta$ is anomalously diffusive in the sense that $H=0$. This is sometimes referred to as \emph{superslow diffusion}.\cite{PhysRevE.66.046129}
In comparison, Ref. ~\onlinecite{PhysRevB.65.104516} found that for the corresponding normal phase of a single resistively shunted Josephson junction, 
the MSD follows the power law \eqref{w_scale} with the exponent decreasing continuously with dissipation strength 
(from $H = 1/2$ for $\alpha \gtrsim 0$ to $H \approx 0$ for $\alpha = \alpha_c$). 

\begin{figure}
  \centering
   \includegraphics[width=0.45\textwidth]{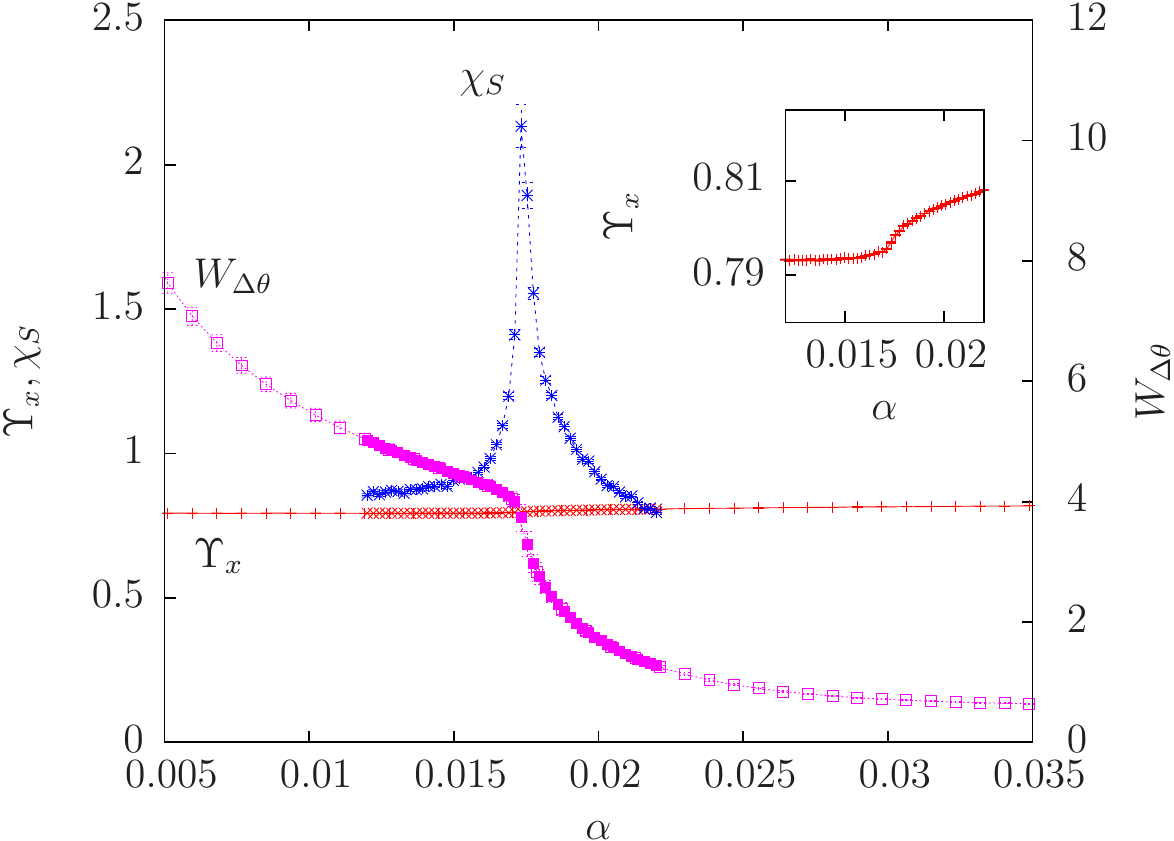}
  \caption{ (Color online) The spatial helicity modulus $\Upsilon_x$, Eq. \eqref{helicity}, action susceptibility $\chi_S$, Eq. \eqref{susceptibility}, and mean fluctuation 
   $W_{\Delta\theta}$, Eq. \eqref{W}, as a function of dissipation strength $\alpha$ for a system in the strong Josephson-coupling regime with $K_\tau=0.002$, $K=1.5$, 
   $N=20$ and $N_\tau = 250$.  Inset: A blow-up of the helicity modulus around the $\alpha_c^{(2)}$ transition. For $\alpha>\alpha_c^{(2)}$, the dissipation renormalizes 
   the spatial coupling strength so that a kink in $\Upsilon_x$ is visible at the localization transition. However, this renormalization is miniscule. Proliferation of 
   instantons across the line $\alpha_c^{(2)}$ in Fig. \ref{fig:phasediag} does  not trigger a proliferation of vortices.}
  \label{fig:div_strong}
\end{figure}

As we show in  Fig. \ref{fig:div_strong}, $W_{\Delta\theta}$ as well as the action susceptibility $\chi_S$ and the helicity modulus $\Upsilon_x$ all 
feature nonanalytic behavior at the critical value $\alpha_c^{(2)}$. Fig. \ref{fig:div_strong} therefore supports the notion that the transition at 
$\alpha_c^{(2)}$ is indeed a genuine dissipation-induced quantum phase transition. Since we have shown that the system for $\alpha > \alpha_c^{(2)}$ has 
both spatial phase coherence and temporal localization of $\Delta \theta$, we can identify this region as a fully bond-ordered superconducting (FSC) 
phase. However, $\Upsilon_x > 0$  even for $\alpha<\alpha_c^{(2)}$, indicating that also the weak-dissipation CSC phase features spatial phase coherence.
The kink in the helicity modulus shown in the inset in Fig. \ref{fig:div_strong} may be attributed to the (slightly) reduced spatial rigidity as the 
bond variables delocalize in imaginary time when leaving the FSC phase. An important conclusion to be drawn from this, is that in the regime of strong Josephson coupling, proliferation of instantons does not trigger a proliferation of vortices at $\alpha_c^{(2)}$ in Fig. \ref{fig:phasediag}. 

A possible physical interpretation of the behavior at strong Josephson coupling and weak dissipation is a phase where there are fluctuations of voltage (and thus also 
of normal currents through the shunts) even though a finite superfluid density allows the system as a whole to sustain an unimpeded supercurrent. For reasons that 
will be apparent in the next section, we have chosen to refer to this state as a critical superconducting (CSC) phase. Similar conclusions have been made earlier 
for (1+1)D systems, \eg, in Refs. \onlinecite{PhysRevB.75.014522}, \onlinecite{PhysRevB.41.4009}, and \onlinecite{PhysRevB.45.2294}, where the authors claimed to have 
found an additional superconducting state characterized by spatial coherence but large local fluctuations. An experimental signature of the FSC--CSC phase transition 
would be to measure an abrupt increase in voltage fluctuations across each junction while the system maintains a Josephson current across the system as the dissipation strength 
is reduced. The phase CSC therefore represents a locally metallic (on each junction) and globally superconducting (throughout the system) state.

\section{The NOR--CSC transition}\label{sec:BKT}

We next consider the phase transition separating the CSC phase from the fully disordered state NOR. First, we {note} that the region at weak dissipation and low 
Josephson coupling in the phase diagram of Fig. \ref{fig:phasediag} is spatially phase incoherent, $\Upsilon_x = 0$. This is therefore identified as the normal, 
metallic phase (NOR) of the dissipative JJA. To verify that the CSC state identified in the previous section by its finite spatial coherence is a distinct phase, 
we next show that it is separated from the NOR phase by a genuine phase transition and not just a crossover caused by the  limited spatial extent of the systems.

\begin{figure}
  \centering
   \includegraphics[width=0.45\textwidth]{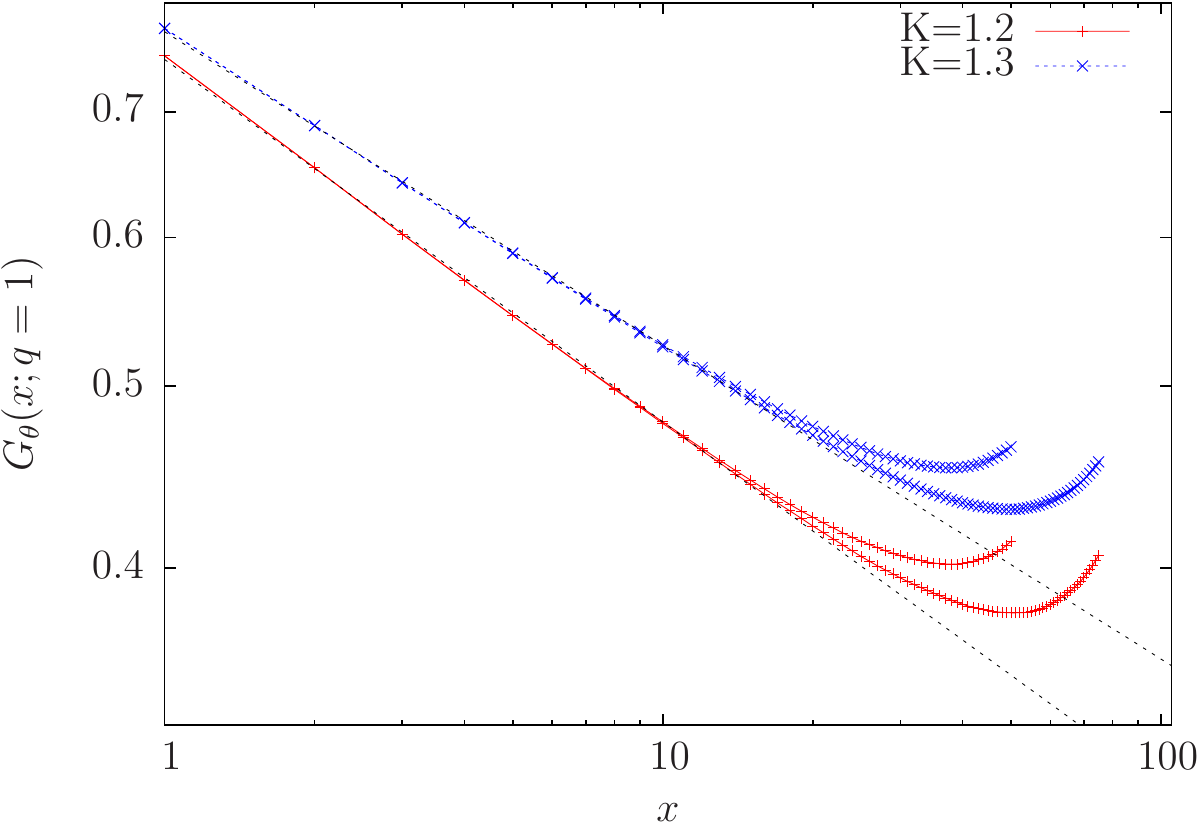}
  \caption{ (Color online) The spatial correlation functions $G_\theta(x;q=1)$, Eq. \eqref{eq:corrq}, calculated for $\alpha=0.005$, $K_\tau = 0.002$, $N_\tau = 20,$ and for two values of the Josephson 
coupling and two values of the spatial extent $N$. Both coupling values correspond to the CSC phase. The dotted lines show the power-law fit of the correlation functions.  }
  \label{fig:corr_strong_power}
\end{figure}

In Fig. \ref{fig:corr_strong_power}, we show algebraically decaying correlation functions in the spatial direction in the CSC phase, indicating QLRO within 
each Trotter slice. In combination with the observation of vanishing order in the temporal direction (as measured by $W_{\Delta\theta}^2$), this motivates an interpretation 
of the CSC phase as a dimensionally reduced critical phase in which the Trotter slices are decoupled from each other. We verified that varying $N_\tau$ had no impact on the 
results for any of the observables probing spatial behavior. Thus, the extent of the systems is fixed at $N_\tau=20$ in the following. 

We anticipate the phase transition separating the NOR phase from the CSC phase to be in the BKT universality class. At the transition point, the helicity modulus is 
expected to scale according to the finite-size scaling function\cite{PhysRevB.37.5986}
\begin{align}\label{eq:BKT}
\Upsilon_x(N) = \Upsilon_x(\infty)\left( 1 + \frac{1}{2}\frac{1}{\ln{N} + C}   \right),
\end{align}   
where $\Upsilon_x(\infty)$ is the value of the helicity modulus as $N\rightarrow \infty$ and $C$ is an undetermined constant. The critical value $K_c$ may be extracted by 
varying $K$ until an optimal fit is achieved. In addition, at a BKT transition, the value of $\Upsilon_x(\infty)$ obtained at optimal fit {should satisfy the universal 
relation} $\Upsilon_x(\infty) K_c = 2/\pi$.

By treating both parameters as variables in the fitting procedure, no \emph{a priori} assumption on the value of the jump is made. This value may consequently 
be used as an additional check on the validity of the conjecture of identifying the transition as a BKT transition.

\begin{figure}
  \centering
    \subfigure[The spatial helicity modulus $\Upsilon_x$ as a function of spatial system sizes $N$ for $\alpha=0.0$, $K_\tau=0.002$ and various Josephson coupling 
    values.]{\label{fig:heliscale_a}\includegraphics[width=0.45\textwidth]{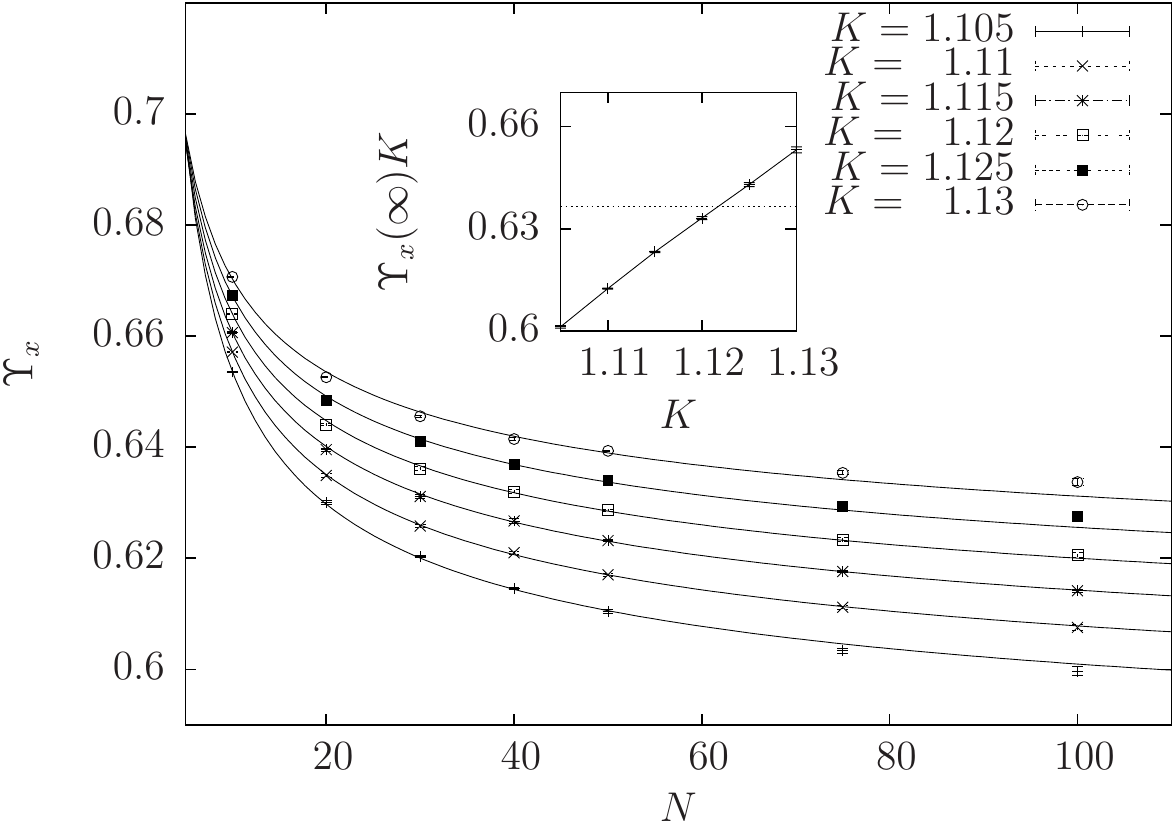}}
   \subfigure[Spatial helicity modulus $\Upsilon_x$ as a function of spatial system sizes $N$ for $\alpha=0.005$, $K_\tau=0.002$ and various Josephson coupling values.]{\label{fig:heliscale_b} \includegraphics[width=0.45\textwidth]{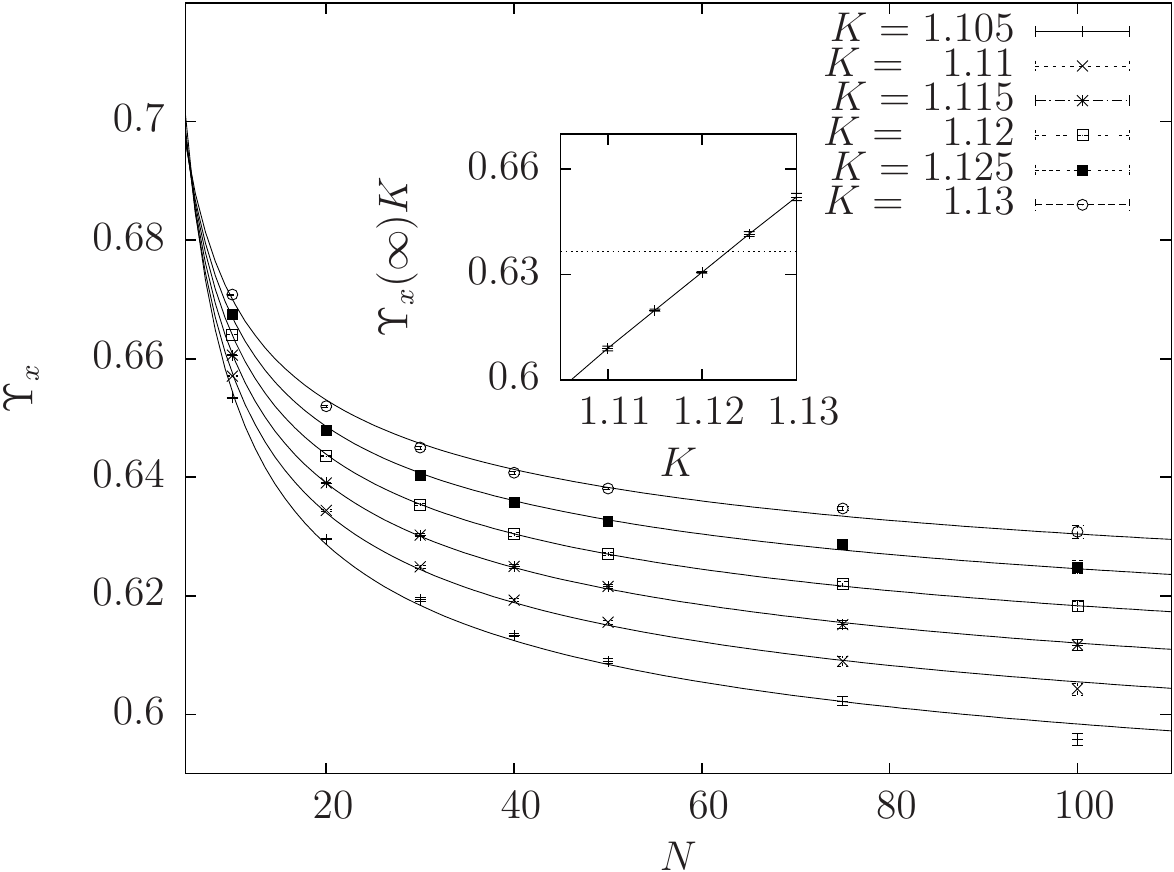}}
  \caption{Comparison of calculated values of the spatial helicity modulus $\Upsilon_x$ with the scaling function \eqref{eq:BKT} for two different dissipation strengths. 
For both values of $\alpha$, a good fit is observed at $K \approx 1.12$. Inset: The universal jump of the helicity modulus is expected to be $2/\pi$ for a BKT transition. 
This value is indicated by a broken line in the insets. The universal jump as calculated from the fitting procedure is shown to be in good correspondence with the BKT 
scenario.}
\label{fig:heliscale}
\end{figure}

In Fig. \ref{fig:heliscale} we present $\Upsilon_x$ for various spatial system sizes and the corresponding fit with Eq. \eqref{eq:BKT}. Fig. \ref{fig:heliscale_a} shows 
results for the dissipationless limit, $\alpha=0$,\cite{footnote_previously47}
while Fig. \ref{fig:heliscale_b} gives the corresponding results for $\alpha=0.005$. At both dissipation strengths 
we observe optimal fit at $K \approx 1.12$. The insets presented in both figures show $\Upsilon_x(\infty)K$, which should be compared to the broken line indicating 
the expected $2/\pi$ universal jump of a BKT transition. These results demonstrate that the NOR--CSC transition is a BKT transition. The temporal interaction 
terms are evidently completely incapable of establishing temporal order at this transition. In particular, when comparing Fig. \ref{fig:heliscale_a} and Fig. \ref{fig:heliscale_b}
corresponding to no dissipation and weak dissipation, respectively, no significant difference is visible. Even though the dissipation term has a major impact on the temporal 
fluctuations,\cite{footnote_previously48}
the spatial helicity modulus appears completely unaffected by the presence of dissipation in the CSC phase.

The classification of the NOR--CSC transition is important in two respects. Firstly, the finite-size analysis shows that the existence of a finite 
helicity modulus in the CSC phase is not a mere finite-size effect. Secondly, the analysis places the transition in the BKT universality class. This 
would not have been possible if there were a divergent correlation length in the temporal direction. Such an effect would have been likely to show up 
as a break-down of the scaling procedure. In this way the analysis gives an indirect verification that the transition is of a purely spatial nature,
and that the CSC phase is temporally disordered and spatially quasi ordered.

\section{The NOR--FSC transition $\alpha_c^{(1)}$}
\label{sec:results_low}

The transition line $\alpha_c^{(1)}$ is the only transition line in the phase diagram that exhibits a simultaneous temporal and spatial 
order-disorder transition. Hence, it involves an interplay between instantons (or instanton-like objects) and vortices, but in a complicated 
way that is not easy to disentangle. 

In this section, the Josephson coupling strength will be fixed at an intermediate value of $K=0.4$, for which a classical counterpart of our model 
would be well inside the disordered phase ($\Upsilon_x = 0)$. The quantum coupling is set to $K_\tau=0.1$. Note that this differs from the value of 
$K_\tau$ used to compute the phase diagram in Fig. \ref{fig:phasediag}.

\begin{figure}
  \centering
   \includegraphics[width=0.45\textwidth]{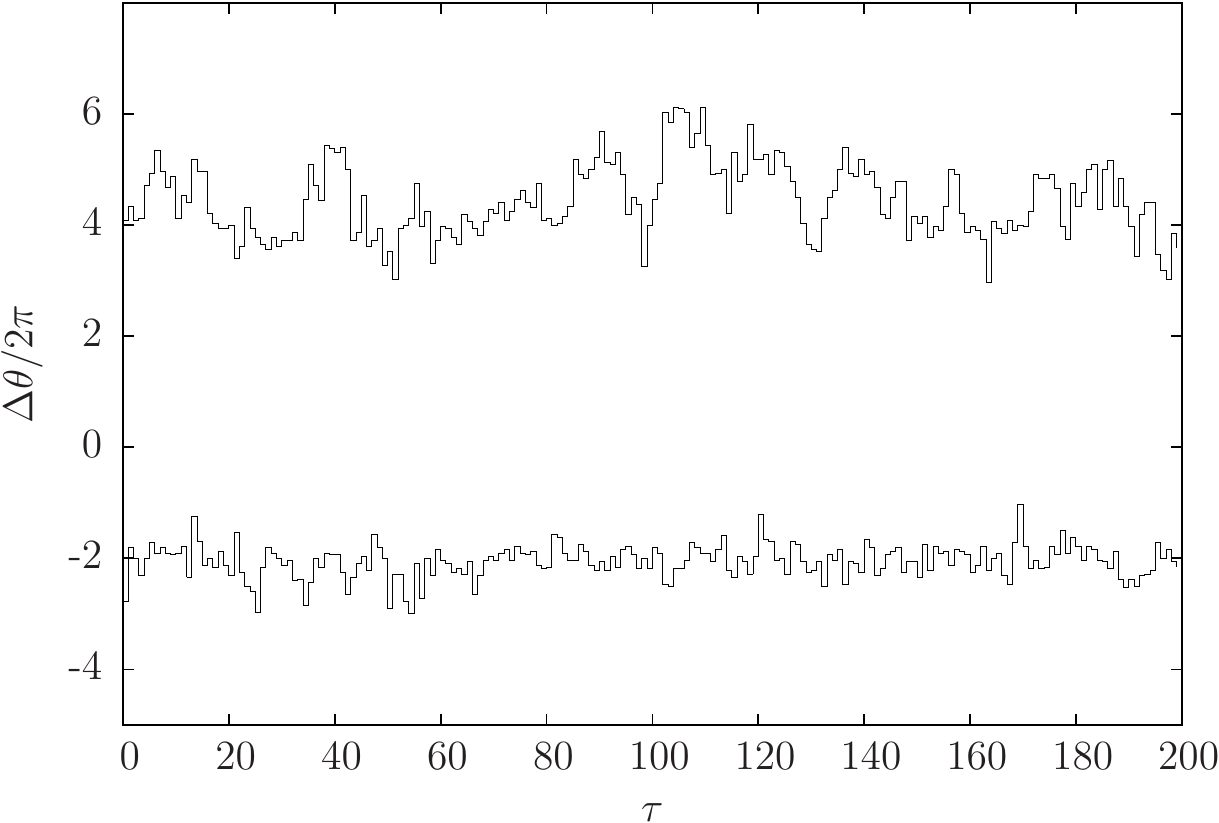}
  \caption{The bond variable $\Delta\theta$ as a function of imaginary time $\tau$ for two different values of the dissipation strength, 
     $\alpha = 0.011$ and $\alpha =0.021$, in the weak and intermediate Josephson coupling regime. These values correspond to 
     the normal phase and the ordered phase, respectively. The 
           quantum paths of $\Delta\theta$ in the normal phase (relatively low values of $K$) exhibit fairly slow variations in time, and are not 
           necessarily well described in terms of instantons. Note the contrast to the quantum paths in the topmost curve above and
           the topmost curve in Fig. \ref{fig:states_strong}.}
  \label{fig:BondRough}
\end{figure}

Fig. \ref{fig:BondRough}  shows typical 
configurations of the bond variable $\Delta\theta$ as a function of $\tau$ for two dissipation strengths corresponding to regimes where the model 
behaves quantitatively different. 
The topmost curve corresponds to weak dissipation, with anomalous diffusive behavior of the 
value of $\Delta\theta$. 
The lowest curve represents the regime of strong dissipation, where the imaginary time-history of $\Delta\theta$ is qualitatively less rough and where 
we can therefore show below that the bond variable is localized.

\begin{figure}
  \centering
   \includegraphics[width=0.45\textwidth]{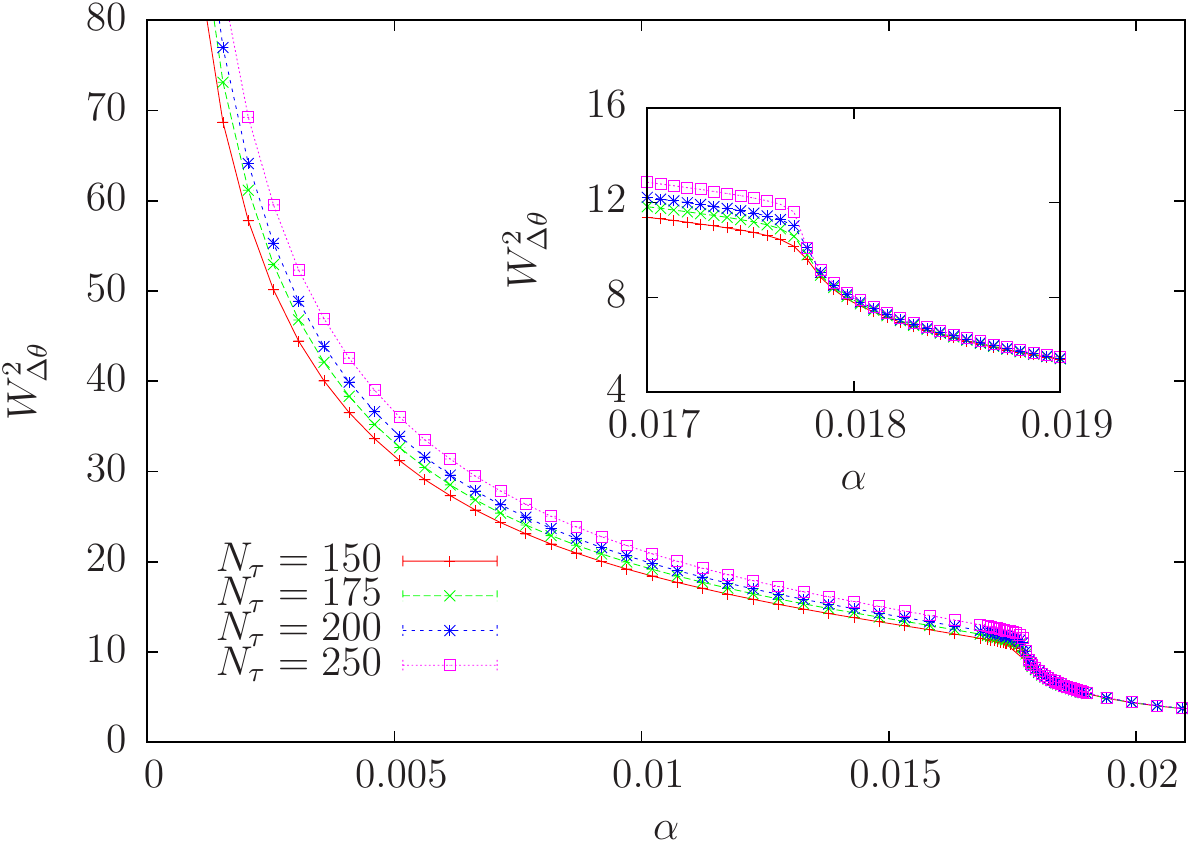}
  \caption{(Color online) $W_{\Delta\theta}^2$, Eq. \eqref{W}, as a function of dissipation strength $\alpha$ for a system with $K_\tau =0.1$, $K=0.4$, $N=20$ 
  and various values of $N_\tau$. Note the kink in the curves at $\alpha=\alpha_c^{(1)}$ and the saturation of $W_{\Delta\theta}^2(N_\tau)$ at a finite value for 
  $\alpha> \alpha_c^{(1)}$. Error bars are smaller than the data points.} Inset: Blow-up of the region around $\alpha_c^{(1)}$. 
  \label{fig:BondFluctVar}
\end{figure}

\begin{figure}
  \centering
   \includegraphics[width=0.45\textwidth]{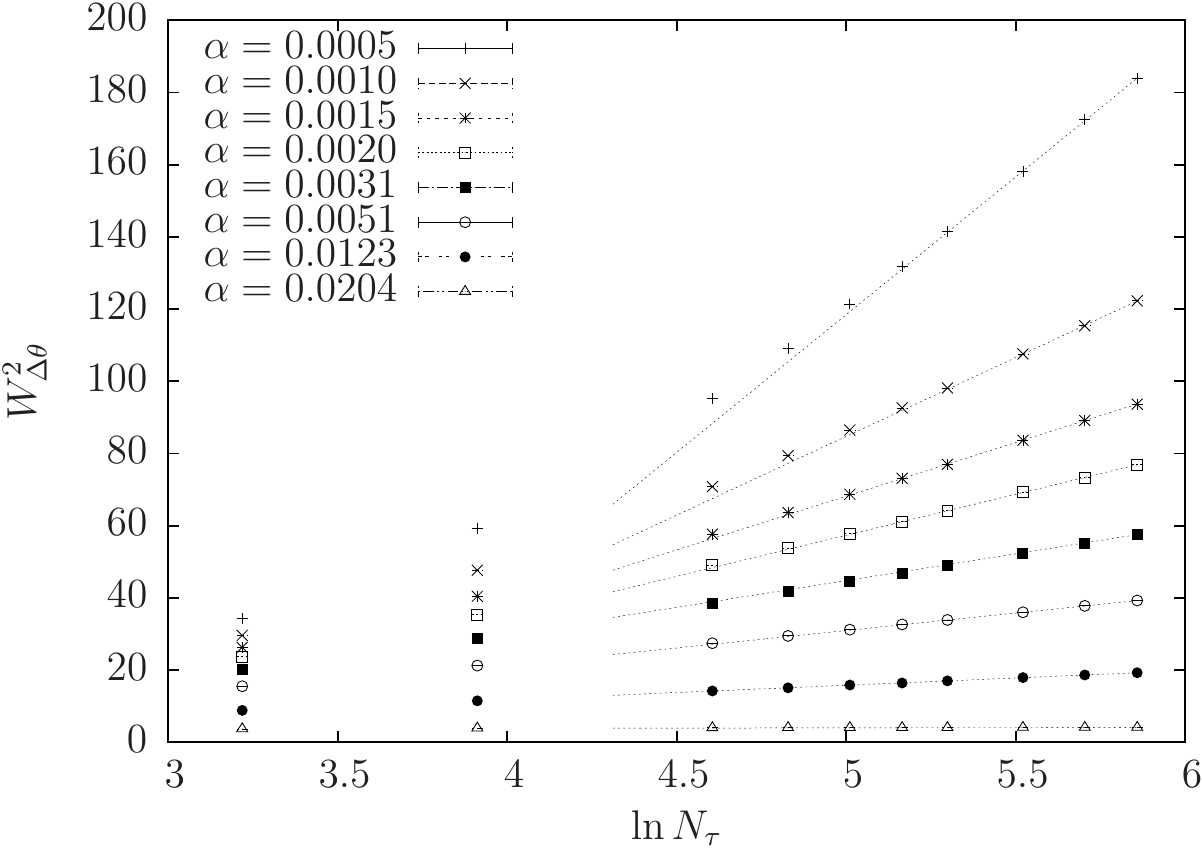}
  \caption{$W^2_{\Delta\theta}$, Eq. \eqref{W}, as a function of $\ln N_\tau$ for various values of $\alpha$ ranging from the weak-dissipation limit to the ordered state at the top and bottom, respectively. The logarithmic behavior found at large $N_\tau$ is indicated by dotted lines. Error bars are much smaller than the data points.} 
  \label{fig:H_alpha}
\end{figure}

We see from Fig. \ref{fig:BondFluctVar} that the amplitude of the temporal bond fluctuations 
are rapidly decreasing with increasing $\alpha$. At a critical value of the dissipation strength $\alpha = \alpha_c^{(1)}$, the MSD features a steep drop 
marking the localization transition where the tunneling of $\Delta\theta$ is suppressed sufficiently to give the bond variables a well-defined value in imaginary time.\cite{footnote_previously49}
We are once again able to distinguish between two separate states based on the scaling properties of the MSD. In Fig. \ref{fig:H_alpha}, we present a plot of the MSD as a 
function of $\ln N_\tau$ for several values of $\alpha$. The lowermost curve in the figure again represents the FSC phase, $\alpha > \alpha_c^{(1)}$, where the MSD is 
independent of $N_\tau$. All other curves represent dissipation strengths below the localization transition, and for these a clear logarithmic scaling is observed. In 
this way, there are distinct delocalized and localized regimes for the bond variable also at weak Josephson coupling, and the temporal fluctuations in each of them 
behave in exactly the same way as for strong Josephson coupling.\cite{footnote_previously50}

\begin{figure}
\centering
\includegraphics[width=0.45\textwidth]{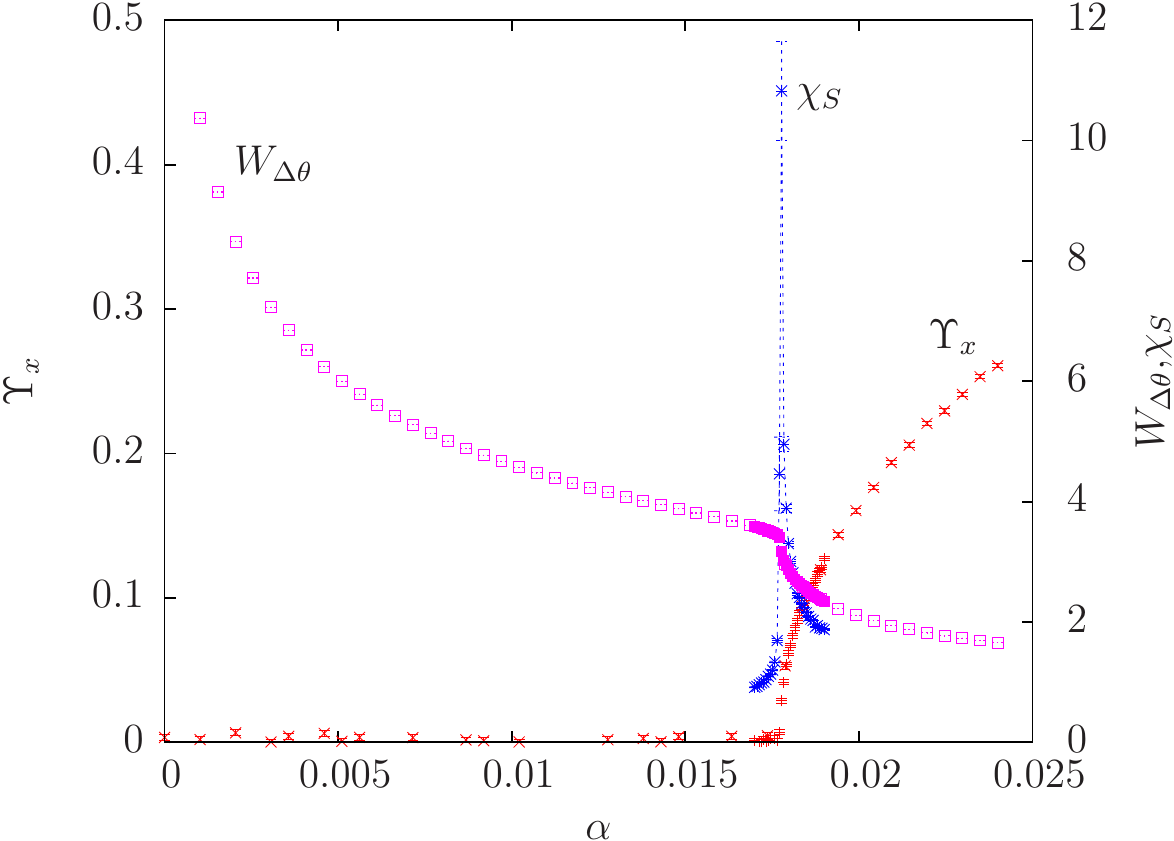}
\caption{(Color online) Spatial helicity modulus $\Upsilon_x$, Eq. \eqref{helicity}, action susceptibility $\chi_S$, Eq. \eqref{susceptibility}, 
                        and mean fluctuation $W_{\Delta\theta}$, Eq. \eqref{W}, as function of dissipation strengths for a system with $K_\tau=0.1$, 
                        $K=0.4$, $N=20$, and $N_\tau = 250$. Note that $\Upsilon_x$ vanishes continuously at $\alpha_c^{(1)}$ (no jump). 
                          }
\label{Fig:div}
\end{figure}

To confirm that the temporal transition at $\alpha =  \alpha_c^{(1)}$ also marks the onset of spatial ordering, we show in Fig. \ref{Fig:div} the helicity modulus 
$\Upsilon_x$. Note the abrupt manner in which the phase stiffness attains a finite value at $\alpha=\alpha_c^{(1)}$. Even though the spatial extent of the system 
is relatively small, there is no weak-dissipation tail which would have been visible for too small system sizes. In the same figure we also show the root mean 
square displacement, $W_{\Delta\theta}$, and the action susceptibility, $\chi_S$. It is clear that all observables feature a nonanalyticity at the same point. 
We can therefore conclude that the transition NOR--FSC is a quantum phase transition involving simultaneous onset of spatial and temporal order.

In Fig. \ref{Fig:div}, we note that the non-analyticity in $\Upsilon_x$ on the line $\alpha_c^{(1)}$ is brought out very sharply at the system sizes we 
consider in this case, namely 20 $\times$ 20 $\times$ 250. Assuming hyperscaling and two diverging length scales $\xi$ (spatial) and $\xi_\tau$ (temporal), we may 
write 
\begin{eqnarray}
\Upsilon_x \sim \xi^{2-d-z} \sim \xi^{2-d} \xi_\tau^{-1} \sim  N_\tau^{-1}.
\end{eqnarray} 
Here we have introduced the dynamical critical exponent $z$ defined by $\xi_\tau \sim \xi^z$.
The sharpness can thus be explained by the large system size and diverging length scale in the $\tau$ direction. Very little finite-size effects may then be expected due to the limited spatial extent of the system, since $d=2$ and the spatial correlation length drops out of the scaling.

Ordinarily, it would have been natural to attempt a scaling analysis of this phase transition based on the Binder ratio
\begin{align}
Q=\frac{\langle|m|^4\rangle}{\langle|m|^2\rangle^2},
\end{align}
in order to extract the dynamical critical exponent of the system, $z$. Here $m$ is the magnetization order parameter of the superconducting  
phases defined in Eq. \eqref{eq:magn}. An ordinary quantum critical point is characterized by diverging lengths in space and time, $\xi$ and 
$\xi_\tau$, respectively. 
The Binder ratio is then expected to 
scale according to
\begin{align}
Q = Q\left(\frac{N}{\xi},\frac{N_\tau}{\xi_\tau}\right).
\label{binder_cumulant}
\end{align}
The correlation lengths entering here are correlation lengths of the phase-correlation function, measuring
$\theta$ correlations in the spatial and $\tau$ directions.
Thus, it should be possible to collapse the Binder ratio curves, at criticality, as a function of $N_\tau/N^z$ for the correct 
value of $z$. 

We have attempted such an analysis in this case, and failed.  In our computations, we have been able to identify
a diverging length scale $\xi$ based on the above scaling approach, but not a diverging length scale $\xi_\tau$.
The reason is that in our model, the coupling in spatial directions is effective in ordering the phases $\theta$,
while the coupling in the $\tau$ direction is only effective in ordering bond variables $\Delta \theta$, while the
$\theta$ variables never order in the $\tau$ direction. One may therefore define a diverging length $\xi$ entering
Eq. \ref{binder_cumulant}, but not a diverging length $\xi_\tau$. A diverging length scale in the $\tau$ direction 
may very well exist for the bond variables $\Delta \theta$, but not for the phase variables $\theta$.

\begin{figure}
\centering
\includegraphics[width=0.45\textwidth]{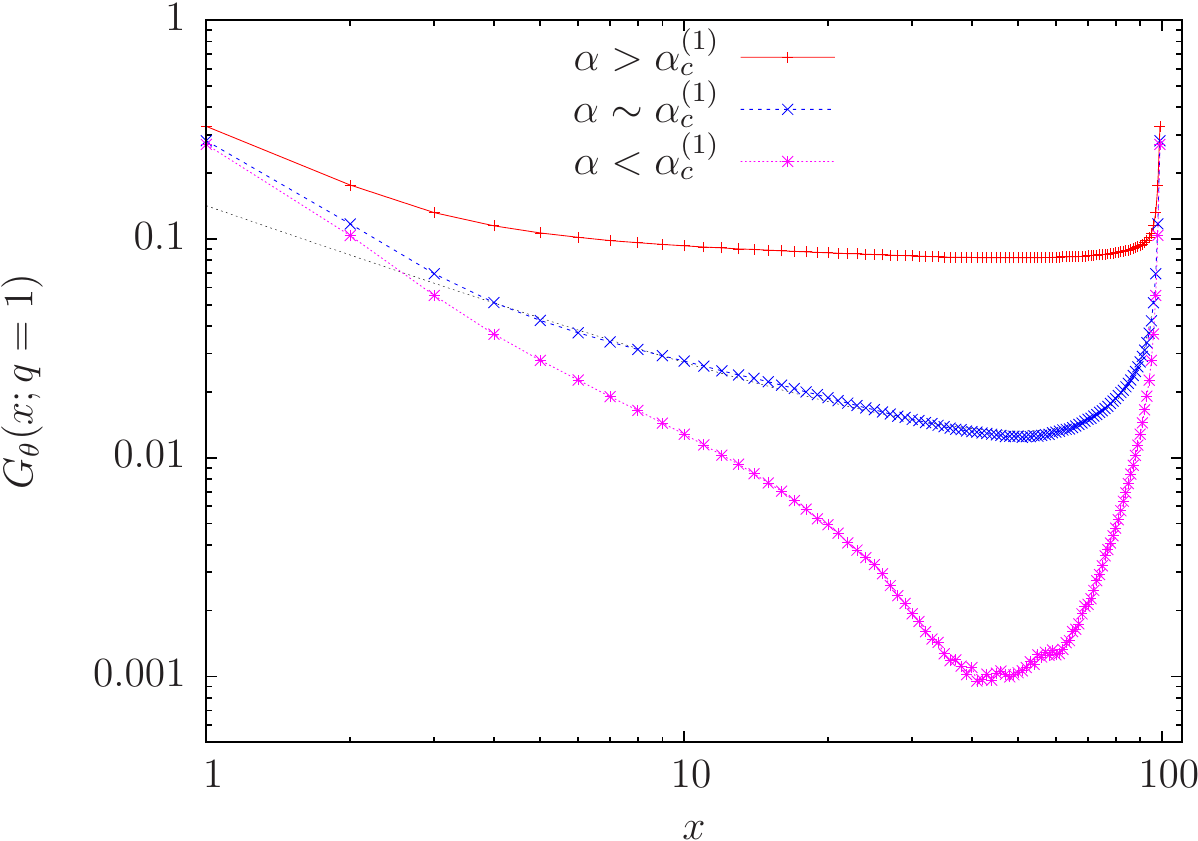}
\caption{(Color online) Double-logarithmic plot of the spatial correlation functions $G_\theta(x;q=1)$ at values of $\alpha$ corresponding to above, close to, and below the NOR--FSC phase transition. The relevant coupling values are $K=0.4$ and $K_\tau=0.1$, and the system size is given by $N=100$ and $N_\tau=30$. At $\alpha \sim \alpha_c^{(1)}$ the correlation function appear to be linear at long distances indicating scale invariant spatial fluctuations. The dotted line indicates this linear behavior. Thus, $\xi \rightarrow \infty$, or at the very least $\xi > N/2$.}
\label{Fig:Corr_NOR_FSC}
\end{figure}

The onset of long-range order in the $\theta$ variables in the spatial directions may be described by the spatial correlation function $G_{\theta}(x;q=1)$, Eq. \eqref{eq:corrq}. In Fig. \ref{Fig:Corr_NOR_FSC} we present spatial correlations corresponding to dissipation strengths slightly below the NOR--FSC transition ($\alpha< \alpha_c^{(1)}$), close to the transition ($\alpha \approx \alpha_c^{(1)}$), and slightly above the transition ($\alpha>\alpha_c^{(1)}$). 
The spatial correlation length appears to behave as expected for a second order phase transition 
into a phase with long-range (spatial) order, implying that the NOR--FSC transition is associated with a diverging length scale in the spatial directions.

%Alternativ caption for log-log plot
%\caption{(Color online) Double-logarithmic plot of the spatial correlation functions $G_\theta(x;q=1)$ at values of $\alpha$ corresponding to above, close to, and below the NOR--FSC phase transition. The relevant coupling values are $K=0.4$ and $K_\tau=0.1$, and the system size is given by $N=100$ and $N_\tau=30$. At $\alpha \sim \alpha_c^{(1)}$ the correlation function appear to be linear at long distances indicating scale invariant spatial fluctuations. Thus, $\xi \rightarrow \infty$, or at the very least $\xi > N/2$.

%\caption{(Color online) Spatial correlation functions $G_\theta(x;q=1)$ at values of $\alpha$ corresponding to above, close to, and below the NOR--FSC phase transition. The relevant coupling values are $K=0.4$ and $K_\tau=0.1$, and the system size is given by $N=100$ and $N_\tau=30$.}

\section{Discussion}\label{sec:discussion}

Since the work of Hertz,\cite{Hertz_quantum_critical} quantum critical points are commonly characterized by their dynamical critical exponent 
$z$. Underlying Hertz' scaling theory is Landau's notion that all relevant fluctuations of a system may be ascribed to fluctuations of an 
order parameter.\cite{PhysRevB.65.165112} This is evident when considering that the exponent $z$ is defined from a divergence of a length scale
of the order parameter correlation function. Such a characterization may therefore be insufficient when the critical point cannot be well 
described by one single order parameter, a problem which has been pointed out in different cases in recent theoretical 
works.\cite{Nature_Si_Local_critical,RevModPhys.77.579} 

The model studied in this paper may be related to a problem of this kind in the sense that we are unable to find one single order parameter adequately 
describing the spatial, temporal, and spatio-temporal phase transitions separating the NOR, FSC, and CSC phases in Fig. \ref{fig:phasediag}. To substantiate 
this, we show in App. \ref{sec:noncompact} that the noncompact $\theta$ variables may instead be formulated by a combination of a compact 
phase field $\tilde{\theta}\in [-\pi,\pi\rangle$ and an additional integer valued field $k$ containing information on what $2 \pi$ interval the original 
variable belongs to.  Using the reformulation of the $\theta$ variables described in Appendix \ref{sec:noncompact}, it is clear that the magnetization 
order parameter $m$ only probes the order of the compactified part of the phase, $\tilde{\theta}$, but is completely oblivious to the state of the 
integer-valued field $k$. Since the state of this field describes whether or not the phase differences $\Delta \theta$ are localized, $m$ is fundamentally 
incapable of describing the localization transition concurring with the onset of coherence of $\tilde{\theta}$. As a result, we 
are unable to define a dynamical critical exponent $z$.

The phase transition  from CSC to FSC is primarily temporal in the sense that it only involves condensation of instantons from a state where 
the spatial topological defects are already tightly bound. However, this localization of $\Delta\theta$ also contributes to spatial ordering by coupling the Trotter 
slices along imaginary time, thereby reducing spatial fluctuations sufficiently to render the system behavior 3D. Accordingly, CSC--FSC is also of a mixed character, 
as the transition separates a phase with spatial QLRO (CSC) from a phase where spatial long-range order is established (FSC). 

The phase transition from NOR to CSC is of a purely spatial nature. As one increases the Josephson coupling for weak dissipation, this transition 
involves only the binding of the (spatial) vortex degrees of freedom while the (temporal) instantons remain proliferated. This conclusion is supported by the 
signatures of a BKT-type transition found in Sec. \ref{sec:BKT}. In this way, the system behaves as a stack of decoupled two-dimensional layers in the CSC phase, 
each exhibiting critical fluctuations in the $\tilde{\theta}$ field.

The phase transition from FSC to NOR is much more complicated than the ones from FSC to CSC and from CSC to NOR, and appears to be of a type not previously 
considered in connection with superconductor-metal phase transitions. Since one cannot characterize the anisotropy of the phase transitions quantitatively in terms of an exponent $z$, we resort to more qualitative considerations 
of the spatial and temporal degrees of freedom. In the case of intermediate coupling, one has a concomitant binding of vortices and localization 
of $\Delta\theta$ upon entering the FSC phase from the NOR phase. This corresponds to the ordering of the degrees of freedom relevant to space ($\tilde{\theta})$ 
and time ($k$), respectively. Due to this simultaneity, we characterize the NOR--FSC phase transition as a mixed spatio-temporal phase transition. It is an 
interplay between two distinct types of topological defects (point-vortices and temporal fluctuations in $\Delta \theta$) that determines the character of the phase transition. This phase transition is therefore neither of the BKT type, nor in the 3D $XY$ universality class. 
The former is characterized by proliferation 
of point-like vortices in two dimensions, while the latter is characterized by the proliferation of (2+1)-dimensional vortex loops\cite{Kleinert,nguyen,tesanovic}. Dissipation, and the associated 
decompactification of the $\theta$ variables, leads to a disordering of the $\theta$ variables in the imaginary-time direction in all regions of the phase diagram. 
Decompactification essentially chops up the vortex loops into spatial point vortices and instanton-like objects in $\Delta \theta$, thereby destroying the 
Lorentz-invariant physics of vortex-loop proliferation at the quantum phase transition. 

In order to exhaust all combinations of spatial and temporal order/disorder, one could also imagine a fourth phase exhibiting temporal order without accompanying spatial 
phase coherence. This would correspond to $W_{\Delta\theta}^2 = \mathrm{const.}$ and $\Upsilon_x=0$, \ie, a phase with localized bond variables and proliferated 
vortices. The most probable location of such a phase would be at weak spatial coupling and large dissipation strength, corresponding to the lower right corner of 
Fig. \ref{fig:phasediag}. This scenario opens the possibility of a {purely} temporal ordering coinciding with exponentially decaying spatial correlations upon 
entering this hypothetical phase from the NOR phase. Due to this locality, such a transition could be a possible realization of a local quantum critical point 
(``$z=\infty$'') in a spatially extended system. In order to emphasize that the existence of this local phase is only a possibility that we have not actually 
found in our computations, we have drawn a box of solid lines around the specific region in Fig. \ref{fig:phasediag} and indicated possible realizations of the 
phase transitions by dotted lines. Although the existence of such a phase has been conjectured by analytical work
\cite{0295-5075-9-5-003,springerlink:10.1007/BF00683713,Korshunov} and there is numerical work supporting this view,\cite{PhysRevB.45.2294} we find no signatures 
pointing to the existence of such a local phase in any of the parameter sets considered. Rather, our results strongly indicate that a spatial coupling 
is always rendered relevant by a large enough dissipation parameter $\alpha$.\cite{PhysRevB.73.064503} In this way, the localization of $\Delta\theta$ will always 
induce an onset of spatial phase coherence. This is equivalent to saying that instanton-like excitations will always proliferate prior to, or simultaneously with, 
the unbinding of vortices as the strength of dissipation, $\alpha$, is reduced. {Local quantum criticality (in the sense of having temporal critical fluctuations 
coinciding with spatial disorder) would follow from vortices proliferating prior to instantons as the disordered state (NOR) is approached from the fully 
bond-ordered superdonducting  state (FSC) by reducing $\alpha$.}

\begin{figure}
  \centering
   \includegraphics[width=0.45\textwidth]{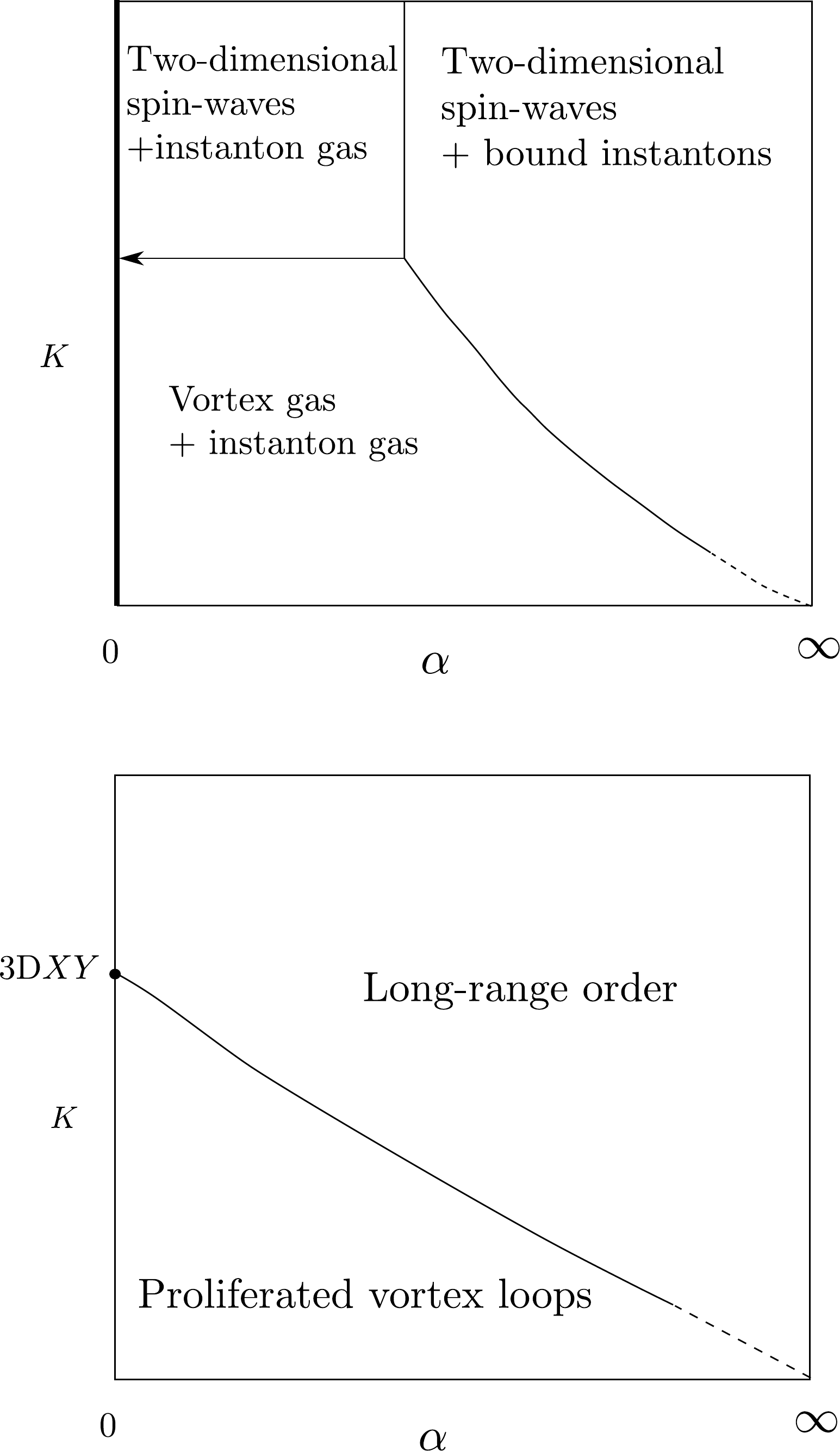}
  \caption{Comparison of the phase diagrams of the noncompact (topmost) and compact (lowermost) models. \emph{Topmost diagram:} The phase diagram found in 
this work. All phases feature disordered $\theta$ variables in the imaginary-time direction. A notable feature is the phase CSC where bound vortex anti-vortex 
pairs coexist with disordered bond variables $\Delta \theta$ in the $\tau$ direction. This 
is a consequence of the $\theta$ variables being defined with noncompact support. This is only true for finite $\alpha$ as the quadratic form of the dissipation 
term is the origin of the decompactification. Consequently, the physics found at finite $\alpha$ cannot be analytically connected to the limit $\alpha=0$.
The description at $\alpha=0$ would require compact phases and thereby a loss of the instanton degrees of freedom. For $\alpha=0$ there is a phase transition at a critical value of $K$, but this phase transition is in the 3D $XY$ universality class, as in the lowermost diagram. $\alpha=0$ is therefore a singular endpoint of the horizontal line 
in the topmost diagram.  \emph{Lowermost diagram:} The phase diagram found for a (2+1)D bond-dissipative quantum rotor model with 
compact variables. In this case the diagram features only a single transition line where the the system undergoes a spatio-temporally isotropic ($z=1$) 
phase transition in the 3D $XY$ universality class.  See Ref. \onlinecite{PhysRevB.84.180503} for details.
     }
  \label{fig:compare_diagram}
\end{figure}

Finally, we compare the phase diagram found in this paper with the phase diagram calculated for a model similar to Eq. \eqref{SBond_Dis} using compact variables.\cite{PhysRevB.84.180503} Fig. \ref{fig:compare_diagram} shows two schematic phase diagrams, and the following discussion pertains to their topology. The topmost diagram summarizes the results found in this paper, while the lowermost diagram is the phase diagram for the $(2+1)$D dissipative quantum rotor model. In the latter case, the diagram features one single phase transition line separating a completely ordered state from a disordered state. The phase transition separating them is driven by a proliferation of vortex loops. This transition line is isotropic in space-time 
($z=1$) meaning that the entire line is in the 3D$XY$ universality class. From the lowermost phase diagram it is clear that if we start in the limit of no dissipation, $\alpha=0$, and increase $\alpha$ for $K>K_{3\mathrm{D}XY}$, the dissipation term only contributes to further dampening the innocuous three-dimensional spin-wave excitations. This can only increase the superfluid density in the ordered phase. However, in the noncompact model the regime $K > K_\mathrm{BKT}$, and weak dissipation, represents a phase involving both two-dimensional spatial spin waves and a proliferated instanton gas. Increasing $\alpha$ from this regime may therefore drive a phase transition because the dissipation term is effective in binding these temporal defects. Therefore, the feature of the phase diagram of the noncompact model that really sets it apart from the phase diagram of the dissipative 3D $XY$ model (i.e., the compact case), is the existence of a phase at weak dissipation involving spatial ordering concomitant with temporal disorder. The resulting phase CSC has no counterpart in the dissipative 3D $XY$ model, since in the latter model the phases $\theta$ are compact. Compact phases $\theta$ promote vortex loops as the critical fluctuations, while noncompact phases $\theta$ promote vortices and instantons as relevant fluctuations driving the phase transitions.   

The phase CSC corresponds to a resistively shunted Josephson junction array which may sustain a finite Josephson current through the array, but nonetheless features finite voltage fluctuations across each junction of the junction array. This may be viewed locally (at a single junction) as a metallic state, but globally (throughout the system) as a superconductor. The most complicated aspect of the phase diagram of the noncompact model is the direct phase transition between the NOR phase 
and the FSC phase, which is considerably more difficult to characterize than the $z=1$ order-disorder transition in the dissipative 3D $XY$ model.

A (2+1)-dimensional model with bond dissipation has recently been considered as an effective theory of quantum criticality at optimal doping in high-$T_c$ 
cuprates.\cite{Aji-Varma_orbital_currents_PRL} The claim of this work is that the phase correlators of the model at the critical point decay algebraically as  $1/\tau$ while they are short-ranged 
in space. Such a phase transition would be an example of local quantum criticality. Monte Carlo simulations on the (2+1)-dimensional quantum rotor model gives an order-disorder 
transition in the 3D $XY$ universality class, which is quite different from local quantum criticality. From the results of the present paper, 
it appears to be important to specify whether the phase variables are compact or noncompact, c.f. Fig. \ref{fig:compare_diagram}.
The phase transitions separating the CSC phase from the FSC phase, or the CSC phase from the NOR phase, are not of the type described in Ref. 
~\onlinecite{Aji-Varma_orbital_currents_PRL}. To verify whether or not the remaining phase transition separating the FSC phase from the NOR phase is an example 
of local quantum criticality one would ideally need a single order parameter measuring spatial and temporal correlations in phases, $\theta$. Since 
we do not have this at our disposal, we have not been able to determine what sort of universality class the critical line separating FSC and NOR belongs 
to, apart form concluding that it is not in the 2D $XY$ or 3D $XY$ universality class. However, the spatial correlation functions presented in Fig. \ref{Fig:Corr_NOR_FSC} suggest that the NOR--FSC transition line is not a line with local spatial phase correlations.

We end with an important remark on the temporal phase fluctuations we have focused on in this paper. The quantity  $W_{\Delta\theta}^2$ in Eq. \ref{W}
measures temporal fluctuation in  {\it{phase gradients}} $\Delta \theta$, defined on a {\it{spatial}} bond of the lattice. One could also study a 
corresponding measure of temporal fluctuations of the phases $\theta$ themselves. We have done this, and find the following. In all parts
of the phase diagram in Fig. \ref{fig:phasediag}, the quantity
\begin{eqnarray}
W_{\theta}^2(N_\tau) = \frac{1}{N_\tau}\left\langle\sum_\tau^{N_\tau} \left(\theta_\tau - \overline{\theta} \right)^2 \right\rangle.
\end{eqnarray}
diverges with $N_\tau$. This underlines that the instantons or instanton-like objects we have discussed in this paper are temporal fluctuations 
in phase gradients $\Delta \theta$, not instantons in phases $\theta$. On the other hand, the helicity modulus Eq. \ref{helicity} 
measures long-range or quasi-long-range spatial ordering of phases $\theta$, and we find such orderings in the FSC and CSC phases. Thus, 
the FSC phase does not exhibit 3D $XY$ ordering. It features spatial ordering of $\theta$ and $\Delta \theta$, but temporal ordering 
only of $\Delta \theta$. This supports the statement made above, that the NOR--FSC transition is not in the 3D $XY$ universality class. 
It is a new type of phase transition involving a complicated interplay between spatial point-vortices and instanton-like
excitations in $\Delta \theta$.   

\section{Conclusions}\label{sec:conclusion}

The model discussed in this paper describes a two-dimensional array of quantum dissipative Josephson junctions. By extensive Monte Carlo simulations we have shown that 
the model features three distinct phases (see Fig. \ref{fig:phasediag}) featuring different behaviors of spatio-temporal fluctuations. {We have quantified these
fluctuations by the mean square fluctuation $W_{\Delta\theta}^2$, Eq. \eqref{W}, and the spatial helicity modulus $\Upsilon_x$, Eq. \eqref{helicity}.}

The normal phase (NOR) is found at weak dissipation and weak Josephson coupling strength. In this phase, the spatial helicity modulus is zero, signaling a vanishing 
stiffness to infinitesimal phase twists on each Trotter slice. The phase differences of the individual junctions are highly fluctuating in imaginary time and the 
system therefore exhibits metallic behavior. Increasing the dissipation strength drives the system to a phase transition where the phase differences $\Delta\theta$ 
are localized into one of the minima of the Josephson potential. This localization of bond variables in imaginary time occurs simultaneously with an onset of rigidity 
towards phase-twists across the spatially extended system. We identify this phase with a fully bond-ordered superconducting state (FSC).

At strong coupling and weak dissipation we identify an intriguing phase exhibiting finite phase stiffness and algebraically decaying spatial correlations. The 
imaginary-time direction remains disordered with wildly fluctuating bond differences. This dimensionally reduced phase is referred to as a critical superconducting 
(CSC) phase. The finite helicity modulus in this phase indicates that the system may sustain a dissipationless current going through the entire JJA. There are, 
however, voltage fluctuations present which in principle should make it experimentally distinguishable from a fully bond-ordered superconducting phase, and also distinct
from the more standard 3D $XY$ ordered fully superconducting state where even the phases $\theta$ are ordered in all directions.

{We have found no signs of a phase which is temporally ordered (in the sense of having a bounded $W^2_{\Delta \theta}$) and proliferated vortices. 
Such a phase would naturally facilitate the  observation of local quantum criticality in which a spatially disordered and temporally (quasi-)ordered 
system disorders in the imaginary-time direction. }

\acknowledgments
The authors acknowledge useful discussions with E. V. Herland, A. Hansen, I. Simonsen, V. Aji, C. M. Varma, and J. Zaanen. A. S. was supported 
by the Norwegian Research Council under Grant No. 205591/V30  (FRINAT). E.B.S. and I.B.S thank NTNU for financial support. The work was also 
supported through the Norwegian consortium for high-performance computing (NOTUR). A. S. thanks the Aspen Center for Physics, where part of this 
work was done, for hospitality.

\appendix
\section{Reformulating the noncompact degrees of freedom}\label{sec:noncompact}

To gain further insight into the three phases reported in this work and the transitions between them, we consider the following decomposition of the phase 
degrees of freedom:
\begin{align}\label{eq:reform}
\theta_{\mathbf{x},\tau} \rightarrow \tilde{\theta}_{\mathbf{x},\tau} + 2\pi k_{\mathbf{x},\tau}. 
\end{align}
The noncompact starting point $\theta$ is thereby exchanged for a compact phase field, $\tilde{\theta} \in [ -\pi, \pi \rangle$, plus an integer-valued 
field, $k$, keeping track of the specific $2\pi$ interval the original variable belonged to. In the partition function, this reformulation amounts to
\begin{align} 
Z =& \int \mathcal{D}\theta \e{-S}  = \int_{-\infty}^\infty \prod_{\mathbf{x},\tau}\left(  \mathrm{d}\theta_{\mathbf{x},\tau} \right)\e{-S} \\ \nonumber
&\rightarrow \sum_{\{ k \}} \int \mathcal{D}\tilde{\theta} \e{-S} = \sum_{\{k\}}\int_{-\pi}^\pi \prod_{\mathbf{x},\tau}\big(\mathrm{d}\tilde{\theta}_{\mathbf{x},\tau} \big)\e{-S}.
\end{align}
Note that $k$ is defined on every point in space-time and has nothing to do with the winding number found in some 
realizations of quantum rotor models with compact phases.

It should also be noted that the $2\pi$-periodic spatial interaction is only sensitive to the $\tilde{\theta}$ field. Also, the compactness of $\tilde{\theta}$ 
enables the identification of vortices in this field in a similar way as discussed in connection with the classical 2D $XY$-model, Eq. \eqref{eq:2DXY}. The finite 
$\Upsilon_x$ observed in the CSC and FSC phases may thereby be attributed to phase coherence in $\tilde{\theta}$. In addition to the vortex degrees of freedom found 
in the classical version of the system, the noncompactness of the quantum version introduces an additional degree of freedom ($k$) associated with the tunneling of 
bond variables from one minimum of the extended Josephson potential to another.

In the NOR phase, we found $\Upsilon_x =0$, which may be understood as a  phase featuring proliferated vortices of the $\tilde{\theta}$ field, as well as proliferated
instantons in $\Delta \theta$. Increasing the Josephson coupling (for small $\alpha$) drives the system into the CSC phase with $\Upsilon_x \neq 0$, which corresponds 
to a binding of vortices into dipoles. Nonetheless, the bond variables remain anomalously diffusive, $W_{\Delta\theta}^2 \propto \ln N_\tau$, in both the NOR and the 
CSC phase. At strong coupling, the bond variables tend to stay in the vicinity of the minima of the potential wells, $\Delta\theta \approx 2\pi\Delta k $.
From the viewpoint of the reformulated variables, the delocalized bond variables in the CSC phase is an expression of an unbroken symmetry 
$\Delta k \rightarrow \Delta k + \Lambda$, where $\Lambda$ is an integer. Moreover, the integer field $k$ may be directly connected with the instanton 
charges in the strong coupling limit by $\Delta k_{\tau+1} - \Delta k_\tau = n_\text{I}$. The delocalization of $\Delta \theta$ manifests itself as 
proliferated instanton/anti-instanton configurations in this regime, and the CSC phase may therefore be described as an instanton gas. This is illustrated in 
the topmost curve of Fig. \ref{fig:states_strong}, illustrating the quantum paths of $\Delta \theta$ in the CSC phase of Fig. \ref{fig:phasediag}. 

For weak Josephson coupling, the excitations in the imaginary-time path of $\Delta \theta$ are strictly speaking not well described by topological instanton 
defects. This is quite evident from the topmost curve of Fig. \ref{fig:BondRough}, describing the quantum paths of $\Delta \theta$ in the NOR phase of 
Fig. \ref{fig:phasediag}. Nevertheless, as the quantum fluctuations still respect the same symmetry $\Delta k \rightarrow \Delta k + \Lambda$, we choose to 
refer to such excitations as instantons also in the NOR phase. In the FSC phase, on the other hand, localization of bond variables implies that the symmetry 
is broken for both weak and strong coupling. Starting at large $K$ and large $\alpha$, the picture is therefore as follows: In the FSC phase, both the defects 
associated with $\tilde{\theta}$ (vortices) and with $k$ (instantons) are absent or tightly bound. Lowering $\alpha$ into the CSC phase, instantons are 
proliferated while the vortices remain bound. Lowering $K$ from the CSC phase into the NOR phase, the vortices proliferate as well.

\section{Correlation functions in the NOR phase}\label{appendix1}

It has recently been proposed\cite{PhysRevB.72.060505,PhysRevB.73.064503} that the metallic state of Josephson junction arrays might exhibit nontrivial behavior. 
Here, it was argued that the (0+1)-dimensional constituents of the array may slide past each 
other in what was denoted a ``floating phase''.\cite{PhysRevB.72.060505} Similar dimensionally decoupled phases are also believed 
to be relevant to other physical systems such as layered superconductors\cite{PhysRevB.45.12632} and stacks of two-dimensional 
arrays of membrane proteins. 

These papers employed a renormalization group analysis to show that the spatial coupling between the superconducting islands is perturbatively 
irrelevant on the disordered side of the transition. 
They also calculated the correlation functions Eq. \eqref{eq:corrq} and \eqref{eq:corrqDelta}  
in this regime and found that they had a form that indicated unconventional, purely local fluctuations.
Monte Carlo studies\cite{PhysRevLett.95.060201} of a single resistively shunted Josephson junction also indicated that a similar form of
correlation functions could be found in (0+1)D systems as well. The correlation functions employed in these analysis featured a noninteger 
parameter $q$ that was introduced to probe fluctuations with another periodicity than the underlying Josephson potential. In the presence of a 
finite Josephson potential, expectation values such as $\langle \e{\mathrm{i}\Delta\theta_{\fat{x},\fat{x'},\tau}} \rangle$ will generally not be 
equal to zero in any phase. This is, however, due to the corresponding symmetry being explicitly -- and not spontaneously -- broken, and 
has consequently nothing to do with a phase transition. The parameter $q$ was therefore introduced to assure correlation functions decaying 
to zero in the disordered phase.
Similar correlation functions have also been considered before in investigations of roughening transitions of crystal surfaces with quenched bulk disorder.\cite{PhysRevB.41.632}
We will refer to them as \emph{fractional correlation functions}.

\begin{figure}
  \centering
   \includegraphics[width=0.45\textwidth]{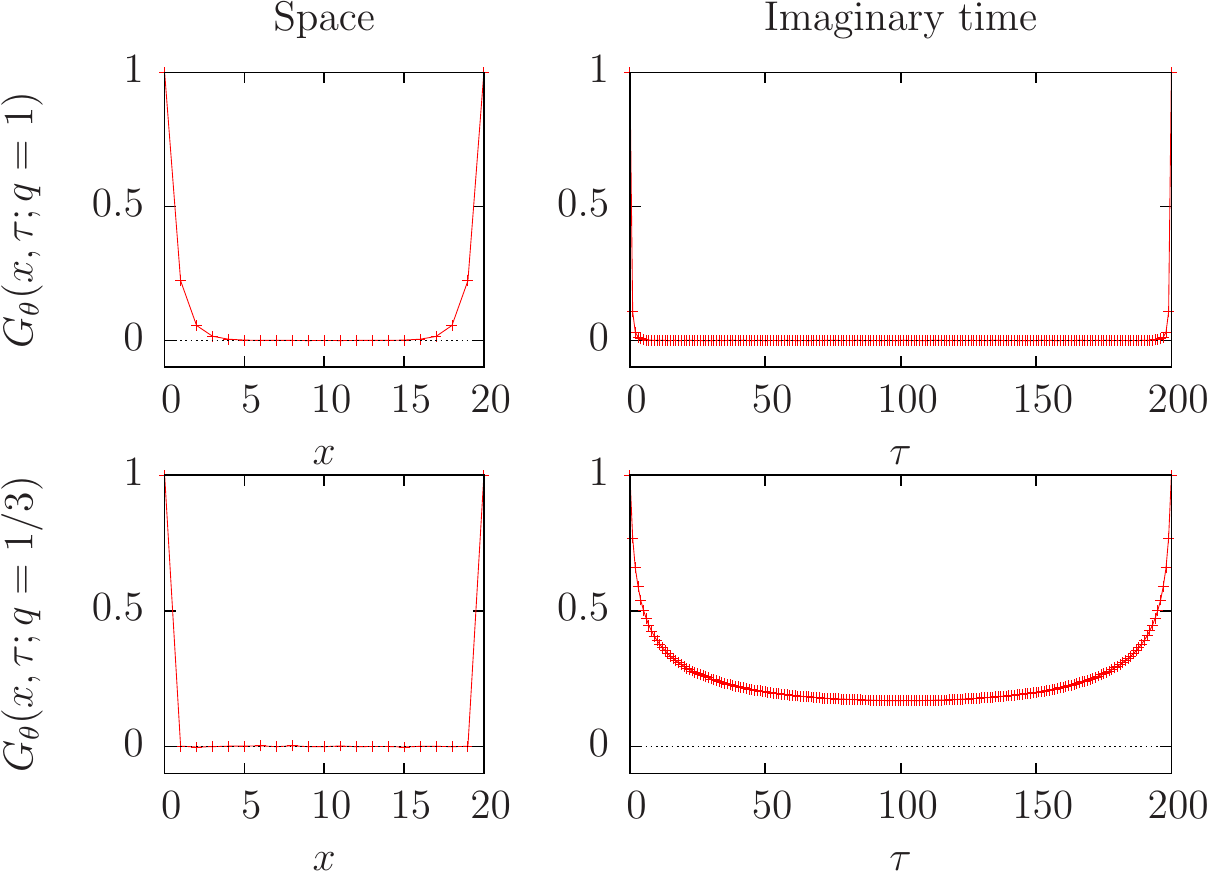} 
    \caption{Correlation functions, Eq. \eqref{eq:corrq}, in both space (left column) and imaginary time (right column) in the normal phase. The relevant 
     coupling parameters are $\alpha = 0.012$, $K=0.4$ and $K_\tau=0.1$.  There is a pronounced difference between the correlations along the spatial and temporal 
     directions for $q=1/3$.}
  \label{Fig:correlations_a}
\end{figure}

Fig. \ref{Fig:correlations_a}, shows both spatial and temporal correlation functions, Eq. \eqref{eq:corrq}, at a dissipation strength deep in 
the NOR phase where the Josephson potential is expected to be irrelevant\cite{PhysRevB.72.060505} and we are far away from the phase transition at 
$\alpha = \alpha_c^{(1)}$. The correlation functions in the bottom row include a noninteger factor $q=1/3$, the top row shows the correlation functions without ($q=1$) this noninteger 
factor. Comparing the correlation function of the temporal direction with the spatial direction for $q=1/3$, it is clear that the spatial and temporal 
behaviors of the system \emph{appear} completely decoupled.\cite{footnote_previously51}
As we discuss below, this local behavior of the fractional correlation functions is misleading.  

App. \ref{sec:noncompact} introduced a reformulation of the phase variables that clarifies the difficulties concerning the construction of a globally order parameter describing our system.  The reformulation of the phase variables also offers an alternative viewpoint on the fractional correlation functions. For 
example, imagine a 2D $XY$ model, Eq. \eqref{eq:2DXY}, being formulated with noncompact phase variables instead of the standard compact variables. In the 
partition function, the summation over $k$ is trivial, yielding only a renormalization of the ground state energy, because there is no coupling between 
different $k$ sectors in the action. The remaining integration over $\tilde{\theta}$ is the partition function of the ordinary 2D $XY$ model. When performing 
Monte Carlo simulations on the 2D $XY$ model with a noncompact formulation of the phases, we find the usual QLRO phase at strong Josephson coupling, in 
which the correlation function $G_{\theta}(\fat{x};q=1)$ of Eq. \eqref{eq:corrq} decays algebraically. However, consider probing the QLRO phase with a 
\emph{fractional} correlation function, $q<1$. This correlation function involves contributions from several $k$ sectors, ultimately averaging the 
correlator to zero for all distances. The same result holds for the disordered phase, and so, although the QLRO phase is phase coherent and the 
disordered phase is not, the fractional correlation function essentially cannot tell them apart.

Applying exactly the same arguments as above to our CSC phase with spatial QLRO, one realizes that the spatial fractional correlation function will vanish also here. 
In analogy with the classical 2D $XY$ model, we argue that this should not be regarded as a signature of completely spatial decoupling in neither the CSC phase nor the NOR phase.
The apparent locality of the normal phase, and by extension the corresponding floating phase of Ref ~\onlinecite{PhysRevB.72.060505}, is consequently not a result of the dissipative interaction \emph{per se}. Rather, a floating phase with such vanishing spatial fractional correlations follows as a direct result of the noncompactness of the phase variables, which in turn is caused by their coupling to a dissipative bath.

\begin{figure}
  \centering
   \includegraphics[width=0.45\textwidth]{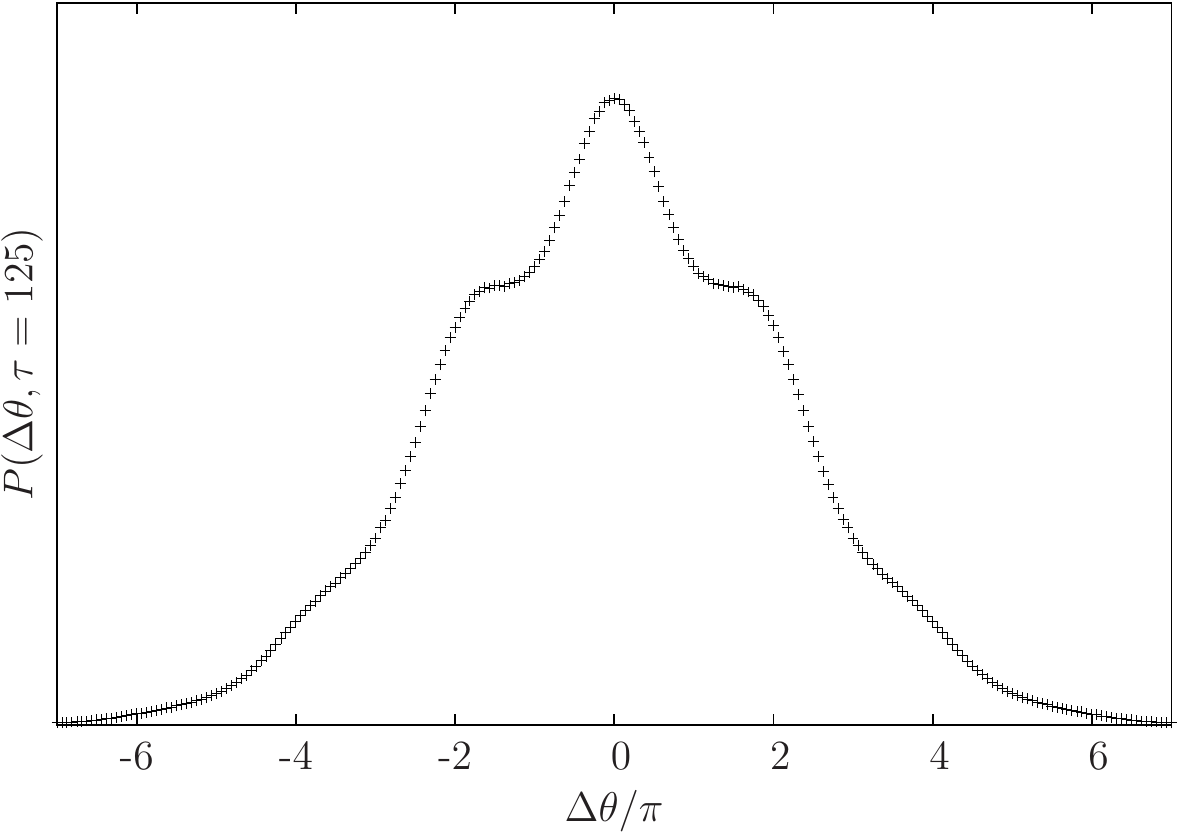}
  \caption{The distribution of $\Delta\theta_\tau - \Delta\theta_0$, $P(\Delta\theta,\tau=125)$, in arbitrary units for a system with $K_\tau=0.1$, $K=0.4$ and dissipation strength $\alpha=0.011$. The distribution is extracted from the Monte Carlo simulations and is conjectured to follow Eq. \eqref{distribution}. }
  \label{fig:bond_dist}
\end{figure}

In the following we provide supplementary details regarding the fractional correlation functions. To be specific, we will investigate the fractional bond correlation 
functions $G_{\Delta\theta}(\tau;q)$ more carefully, and prove that a power-law tail is expected in the weak dissipation regime. We first consider the distribution 
function $P(\Delta\theta,\tau)$, as was also the starting point of Ref. \onlinecite{PhysRevB.73.064503}. This function describes the diffusion of the phase difference 
$\Delta\theta_\tau$ with respect to its value at $\tau = 0$. The distribution broadens for increasing $\tau$ and is illustrated for a fixed imaginary-time distance 
in Fig. \ref{fig:bond_dist}. We find that the distribution function can be very well fitted by the functional form
\begin{align}\label{distribution}
P(\Delta\theta,\tau) = P_0 \e{-\frac{\Delta\theta^2}{2\sigma_G^2}}\sum_n{\e{-\frac{(\Delta\theta-2\pi n)^2}{2\sigma^2}}},
\end{align}
where $P_0$ is a normalization constant.
The distribution is made up of a sequence of sub-gaussians with standard deviation $\sigma$ centered around the minima of the Josephson potential. In addition, there 
is an overall gaussian convolution characterized by a standard deviation $\sigma_G$. We find empirically that whereas $\sigma$ is dependent on $K$, it is independent 
of the distance $\tau$ in imaginary time. The overall variance $\mathcal{G}(\tau)$ of the distribution, as defined by
\begin{align}\label{eq:gradientcorr}
\mathcal{G}(\tau) = \left\langle (\Delta\theta_{\tau} - \Delta\theta_0)^2\right\rangle,
\end{align}
grows logarithmically with $\tau$. This variance can furthermore to a very good approximation be identified with the variance $\sigma_G^2$ of the convolution 
function.

The calculations in Ref. \onlinecite{PhysRevB.73.064503} were based on a strong-coupling limit for the distribution function, with an additional assumption that the spatial 
coupling will renormalize to zero regardless of its bare value. For large values of $K$, we have demonstrated that the system will eventually reach a superconducting 
state, \ie,  the CSC phase or the FSC phase, for all $\alpha > 0$. It is also clear from Fig. \ref{fig:bond_dist} that there is an appreciable broadening of the sub-gaussians 
($\sigma > 0$) compared to the delta-function distribution implicit in the strong-coupling limit ($\sigma \to 0$). 

We next consider the implications of a finite $\sigma$ on the correlation function $G_{\Delta\theta}(\tau;q)$. Assuming Eq. \eqref{distribution}, we calculate
\begin{align}\label{poisson}
\langle \e{\mathrm{i}q(\Delta\theta_\tau - \Delta\theta_0)}\rangle = \e{-\frac{1}{2}\sigma^2 \kappa q^2}\sum_{s=-\infty}^\infty \e{-\frac{1}{2}\sigma_G^2( q-s/\kappa)^2},
\end{align}
where $\kappa = \sigma_G^2/(\sigma_G^2 + \sigma^2)$. The sum over $n$ has been traded for an integral at the cost of introducing an integer Poisson summation variable $s$. 
The $n$-variable is subsequently integrated out. Comparing with Eq. (12) in Ref. \onlinecite{PhysRevB.73.064503}, the broadening of the sub-gaussians has introduced an overall 
prefactor and a multiplicative adjustment of the Poisson summation variable. The strong-coupling result is easily recovered in the limit $\sigma \rightarrow 0$. 
In the limit $\tau \to \infty$, the term with the slowest decay is dominant, hence the sum may be substituted by the term with the smallest $(q-s/\kappa)^2$. For a logarithmically 
diverging $\sigma_G$, we also have $\kappa \rightarrow 1$, meaning that Eq. \eqref{poisson} is a scale-free power law in this limit.

\begin{figure}
  \centering
   \includegraphics[width=0.45\textwidth]{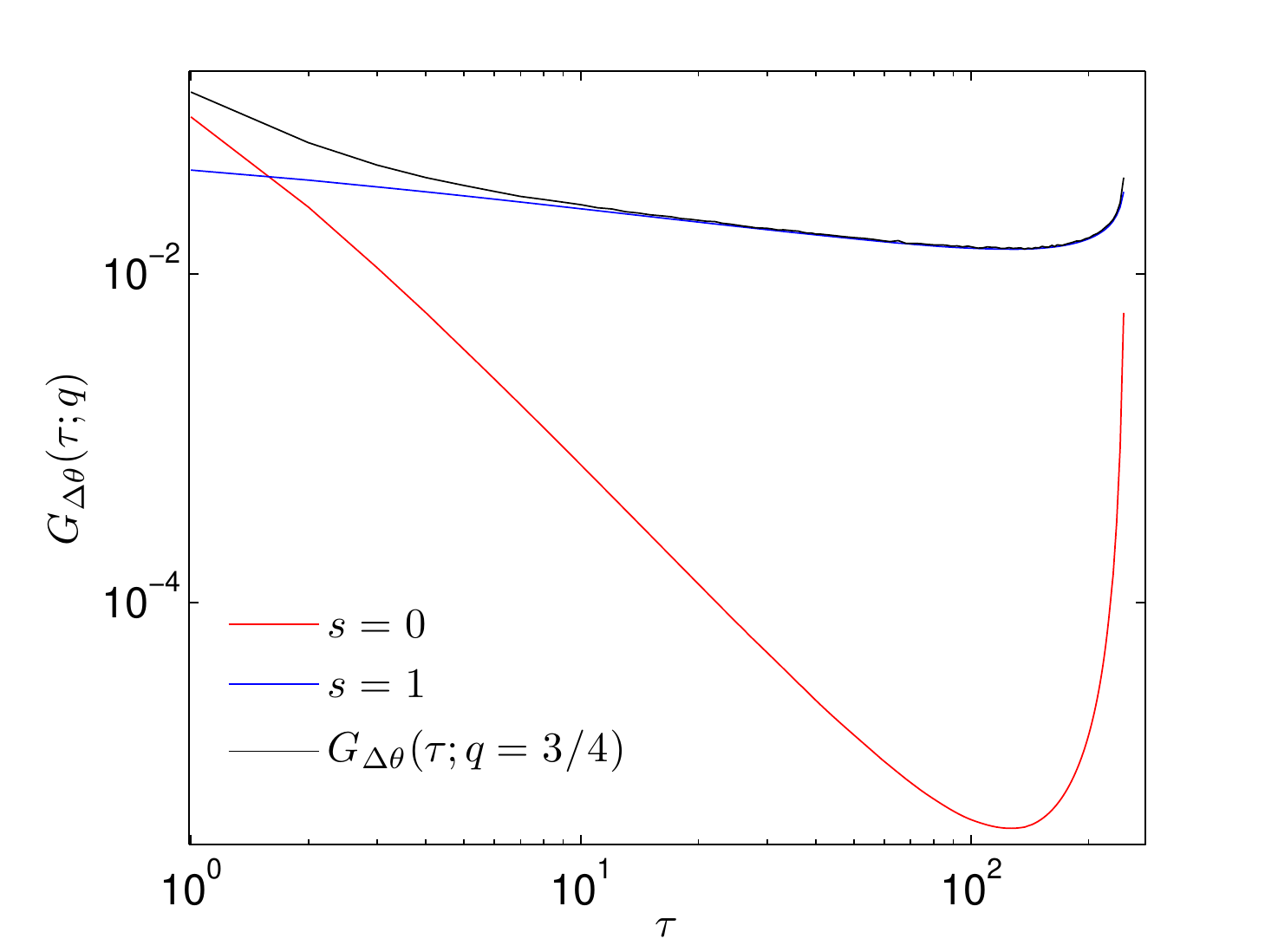}
  \caption{(Color online) The unequal-time bond correlation function, $G_{\Delta\theta}(\tau;q=3/4)$, for a system with $K=0.4$, $K_\tau=0.1$, $\alpha=0.011$, $N=20$ and $N_\tau=250$. The black curve is the correlation function Eq. \eqref{eq:corrqDelta} sampled directly from the Monte Carlo simulations. The red (lowermost gray) and blue (uppermost gray) curve are the $s=0$ and $s=1$ terms of Eq. \eqref{poisson}, respectively, and are calculated as explained in the text. }
  \label{fig:poissoncorr}
\end{figure}

In Fig. \ref{fig:poissoncorr}, we show a plot of $G_{\Delta\theta}(\tau;q=3/4)$ and the two terms from Eq. \eqref{poisson} corresponding to $s=0$ and $s=1$. We have set 
$\sigma_G^2$ equal to $\mathcal{G}(\tau)$ as measured from the Monte Carlo simulations in order to compare the analytical result Eq. \eqref{poisson} with the fractional 
correlation function found numerically. $\sigma$ is specified from fitting Eq. \eqref{distribution} to data from Monte Carlo simulations. At short distances the $s=0$ 
term is still contributing, but a clear cross-over to the dominant $s=1$ term is visible for larger values of $\tau$. The excellent fit between the curves validates the 
functional form of the distribution \eqref{distribution}.

It is interesting to compare the behavior presented above with available numerical results for a single resistively shunted Josephson junction. Ref. \onlinecite{PhysRevLett.95.060201} reports temporal fractional correlation functions in a (0+1)D system that are power law in much the same way as those in Ref. \onlinecite{PhysRevB.73.064503}. They also report a logarithmically diverging MSD, but only at the phase boundary. This is in contrast to the results presented in Secs. \ref{sec:results_low} and \ref{sec:result_strong}, where we find logarithmic growth as a generic feature of the weak-dissipation phases. Following Ref. \onlinecite{PhysRevLett.95.060201}, it is natural to consider the possibility that a logarithmically diverging MSD is the signature of critical behavior for models describing Josephson junctions. A logarithmically diverging MSD follows from a logarithmically diverging $\mathcal{G}(\tau)$, and we have shown that the latter generates fractional bond correlators that are algebraically decaying in imaginary time. A possible scenario could be that the increased dimensionality of the problem has damped the fluctuations such that, in contrast to the single junction, the entire weak-dissipation regime features critical temporal correlations of the bond variables. However, we expect such a critical phase to produce divergent susceptibilities of the action. The simulations do not support this scenario and we find nonanalytic $\chi_S$ only at the points $\alpha = \alpha_c^{(1),(2)}$. Thus, a power-law form of the temporal fractional bond correlators can not necessarily be ascribed to critical behavior of the system.


\begin{thebibliography}{45}
\bibitem[{(Footnote)}]{footnote_previously46}
A dissipation kernel differing from 
the Caldeira-Leggett form has been calculated for coupling of a loop-current order parameter to gapless excitations at the Fermi level of 
high-$T_c$ cuprates. Namely, it was shown in Ref. 
\onlinecite{PhysRevB.77.092404} that the mechanism leading to dissipation is scattering of the bosonic order parameter off gapless fermionic 
particle-hole excitations at the Fermi level. This produces Landau damping of the form $|\omega|/q$, which is a manifestly non-local 
dissipation kernel. For an alternative formulation of the coupling of the bosonic order parameter to gapless fermions on the Fermi surface, 
see Ref. \onlinecite{PhysRevB.81.064515}.
\bibitem[{(Footnote)}]{footnote_previously47}
As mentioned in Sec. \ref{sec:overview}, the limit $\alpha = 0$ is not physical for the model Eq. \eqref{SBond_Dis} with noncompact phase variables. Nevertheless, the spatial behavior of the system changes continuously as one (artificially) reduces the parameter $\alpha$ to zero, so the value $\alpha = 0$ of Fig. \ref{fig:heliscale_a} should simply be regarded as the limit of very weak dissipation.
\bibitem[{(Footnote)}]{footnote_previously48}
Although not explicitly shown, we observed diffusive ($H=1/2$) scaling of $W_{\Delta\theta}^2$ for all values of $K$ in the artificial limit $\alpha=0$. The diffusive behavior of $W^2_{\Delta \theta}$ found at $\alpha=0$, is replaced by logarithmic scaling for any finite value $0 <\alpha < \alpha_c^{(1),(2)}$.
\bibitem[{(Footnote)}]{footnote_previously49}
As a comment relevant to both this section and Sec \ref{sec:result_strong}, a fluctuation measure similar to 
$W^2_{\Delta\theta}$ was calculated in Ref. \onlinecite{PhysRevB.65.104516} for a single dissipative Josephson junction. It was suggested that this quantity exhibited 
a cusp at the localization transition, but later high-quality Monte Carlo simulations revealed a smoothly changing MSD around the critical dissipation strength.
\cite{PhysRevLett.95.060201} Apparently, the $\alpha$ dependence of the MSD is not suitable for locating the critical point in (0+1)D. From the kink shown in the 
inset in Figs. \ref{fig:BondFluctVar} and \ref{fig:rough_strong}, this is not the case for the spatially extended model considered here. A similar kink in a fluctuation measure may also have been present in previous results for a related (2+1)D RSJJA model,\cite{PhysRevLett.94.157001} but this curve was based on too few points for the feature to be convincingly resolved.
\bibitem[{(Footnote)}]{footnote_previously50}
It should be noted that there are stronger finite size effects in $W_{\Delta\theta}^2$ in the NOR 
phase than in the CSC phase.
\bibitem[{(Footnote)}]{footnote_previously51}
Although not explicitly shown here, the fractional bond correlators behave qualitatively similar 
to the site correlators. With a noninteger $q$, $G_{\Delta\theta}(x;q )$ is zero for all spatial distances while $G_{\Delta\theta}(\tau;q )$ is slowly 
decaying in imaginary time.
\end{thebibliography}
\end{document}